\documentclass[useAMS,usenatbib]{mn2e}
\usepackage{times,epsfig,natbib,amssymb,amsmath,graphics,longtable,threeparttable}

\newcommand{\sdssg}{\emph{g$^{\prime}$}}

\def \lsun{\ifmmode{{\rm\ L}_\odot}\else{${\rm\ L}_\odot $}\fi}
\def \msun{\ifmmode{{\rm\ M}_\odot}\else{${\rm\ M}_\odot$}\fi}
\def \rsun{\ifmmode{{\rm\ R}_\odot}\else{${\rm\ R}_\odot$}\fi}
\newcommand{\kms}{kms$^{-1}$}                         % Kms-1
\def \mdot{\ifmmode{{\rm\dot{M}}}\else{${\rm\dot{M}}$}\fi}

\newcommand\as{${''}$}

\newcommand{\ha}{H$\alpha${}}

\newcommand{\NII}{[N{\sc ii}]}                  %[NII]
\title[The radial distribution of CC SNe]{Comparisons of the radial distributions of 
core-collapse supernovae with those of young and old stellar populations\thanks{Based 
on observations made with the Isaac Newton Telescope operated on the island
of La Palma by the Isaac Newton Group in the Spanish Observatorio del Roque de los
Muchachos of the Institute de Astrofisica de Canarias, and 
on observations made with the Liverpool Telescope operated on the island of La Palma by Liverpool John Moores 
University in the Spanish Observatorio del Roque de los Muchachos of the Instituto de Astrofisica 
de Canarias with financial support from the UK Science and Technology Facilities Council.}}
\author[J. P. Anderson and P. A. James]{J. P. Anderson\thanks{E-mail:anderson@das.uchile.cl}$^{1,2}$ and P. A. James$^{2}$\\
$^{1}$Departamento de Astronomia, Universidad de Chile, Camino El Observatorio 1515,
Las Condes, Santiago, Casilla 36-D, Chile\\
$^{2}$Astrophysics Research Institute, Liverpool John Moores University,
Twelve Quays House, Egerton Wharf, Birkenhead, CH41 1LD, UK}
\begin{document}

\date{}

\pagerange{\pageref{firstpage}--\pageref{lastpage}} \pubyear{2008}

\maketitle

\label{firstpage}

\begin{abstract}
We present observational constraints on the nature of core-collapse supernovae through 
an investigation into their radial distributions with respect to those of
young and old stellar populations within their host galaxies,
as traced by \ha\  emission and $R$-band light respectively. We discuss results and the 
implications they have on the nature of supernova progenitors, for a sample of 177 core-collapse
supernovae.\\
We find that the radial positions of the overall core-collapse population closely follow the
radial distribution of \ha\ emission, implying that both are excellent tracers of star formation
within galaxies. Within this overall distribution we find that there is a central deficit of SNII
which is offset by a central excess of SNIb/c. This implies a strong metallicity dependence on the
relative production of the two types, with SNIb/c arising from higher metallicity progenitors than SNII. 
Separating the SNIb/c into individual classes we find that a trend emerges in terms of
progenitor metallicity going from SNII through SNIb to SNIc, with the latter arising from the highest
metallicity progenitors. 
%With respect to the SNII sub-types we find the surprising result that the SNIIP are found nearer to
%the centres of host galaxies that the other sub-types (IIL, IIb, IIn and `impostors'). This implies 
%that the various other SNII sub-types arise from less metal rich progenitors than SNIIP.

\end{abstract}

\begin{keywords}
stars: supernovae: general -- galaxies: general -- galaxies: statistics
\end{keywords}

\section{Introduction}
\label{intro}
Determining the characteristics of supernova (SN) progenitors is a key area of research given 
their importance and influence on many other astrophysical processes. The distribution of SNe within host
galaxies can give substantial information as to the nature of their progenitors. At different
galactocentric radii within galaxies different stellar populations are found with different mean 
ages and metallicities. One can therefore investigate how the radial distributions of SNe correlate
with those of different stellar populations and use this to further constrain the 
nature of progenitor stars. 
%One parameter of SN progenitor stars 
%that is thought to strongly influences the final SN type classification is the initial metallicity.
%Directly determining this metallicity from individual SNe characteristics (light curves, spectra) is not currently possible (IS THIS TRUE?),
%therefore one must use 
%the nature of the stellar populations within which SNe are found to infer metallicities of
%their progenitor stars. One way to investigate this dependence of SN type on progenitor metallicity is to determine the galactocentric
%radial positions where SNe occur, and compare these to metallicity gradients found in galaxies.\\  
Here we present results on the radial distribution of a large number of core-collapse (CC) SNe with
respect to the $R$-(or \textit{r'})-band light and \ha\ emission of their host galaxies,
and discuss the implications these have on the nature of SN progenitors. Results from an initial data set were 
presented in \cite{jam06} (JA06 henceforth) which showed that SNIb/c tended to occur within more central parts of the 
light distribution of their host galaxies, albeit with 
small number statistics (only 8 SNIb/c). This was ascribed to SNIb/c arising from higher metallicity progenitors. 
We test these initial results with an increased sample size enabling us to distinguish 
between the various SN sub-types, and present further results from a combined sample of 115 SNII, 58 SNIb/c
and 4 SN `impostors'. 

\subsection{Core-collapse supernova progenitors}
\label{SNprog}
CC SNe are thought to arise from stars with initial masses $>$8\msun\ (a value that has been converged upon 
from progenitor direct detections: \citealt{sma09}, and 
the maximum observed masses of WDs: \citealt{wil09}). The different CC SN types are then thought
to arise as a result of processes dependent on progenitor mass, metallicity and/or binarity. 
CC SNe are classified according to 
the presence/absence of spectral lines in their early time spectra, plus
the shape of their light curves (see \citealt{fil97}, for a review of
SN classification). The first major classification comes from the presence of
strong hydrogen emission in the SNII. SNIb and Ic lack any detectable
hydrogen emission, while SNIc also lack the helium lines seen in SNIb. SNII
can also be separated into various sub-types. SNIIP and IIL are classified in terms of
the decline shape of their light curves (\citealt{bar79}; plateau in the former and linear in the latter),
thought to indicate different masses of their envelopes prior to SN,
while SNIIn show narrow emission lines within their spectra \citep{sch90}, thought to arise
from interaction of the SN ejecta with a slow-moving circumstellar medium (e.g. \citealt{chug94}).
SNIIb are thought
to be intermediate objects between the SNII and Ib as at early times their
spectra are similar to SNII (prominent hydrogen lines), while at later times they
appear similar to SNIb (\citealt{fil93}).\\
The main advances in our understanding of CC SN progenitors over the past decade have come from the direct detection
of progenitor stars on pre-explosion images. A recent compilation of a decade long search for 
SN progenitors was presented by \cite{sma09}. One of the main conclusions from this work was a
mass range for the production of SNIIP of between 8.5 and 16.5\msun . However, the statistics 
on any of the other CC types currently preclude any firm conclusions as 
to the exact nature of their progenitors. Their spectral and light curve characteristics indicate 
that they must have lost some (the various II sub-types) or
all (SNIb) of their hydrogen envelopes during their
pre-SN evolution, with SNIc additionally having lost all helium. Whether this is due to different mass ranges of progenitors producing stronger
stellar winds and thus removing increasing amounts of mass, or whether metallicity or binarity play a dominant role
in deciding SN type is still under debate.\\
In the previous paper in this series (\citealt{and08}, AJ08 henceforth) we concluded that the main CC SN types
form a sequence of increasing progenitor mass, going from SNII to SNIb and SNIc, with the latter 
arising from the highest mass progenitors. Meanwhile there is
also increasing evidence that SNIb/c tend to arise in more metal rich host galaxies than SNII
(\citealt{pran03}, \citealt{pri08_2}). It therefore seems likely that both progenitor mass and metallicity
play important roles in producing the diversity of observed types. Stellar modeling also
predicts these dependencies (e.g. \citealt{heg03}, \citealt{eld04}) for single star progenitors. There is also a strong likelihood
that massive close binaries produce some of the range of CC SN types. Evidence
for binarity has been claimed for both SN 1987A and SN 1993J (\citealt{pod90} and \citealt{nom93,pod93,maun04} respectively),
while comparisons of the number ratio of SNII to SNIb/c with predictions from single star models and observations of massive stars in 
the local group (\citealt{mass03}, \citealt{crow07}), would seem to suggest that the production of a significant fraction of SNIb/c from binaries
is required.
Therefore while we have yet to tie down the progenitors of the different CC SN types (apart from possibly the SNIIP)
there seem to be many available avenues within our current understanding of stellar evolution to 
produce the observed SN diversity.\\

\subsection{The radial distribution of supernovae}
\label{SNrad}   
At different radial positions within galaxies, stellar populations with different characteristics are thought to dominate. For example
in spiral galaxies the central regions may be dominated by the old stellar bulge while at larger radii younger stellar populations
associated with spiral arms are likely to become relatively more dominant. The metallicities of
particular stellar populations are also observed to strongly correlate with galactic radial position, with higher chemical
abundances found in the centres, decreasing with radial distance. This characteristic is found in nearly all galaxy types with
the steepest abundance gradients found in normal spirals (see \citealt{hen99}, for a review on this subject). One can therefore
study the radial distributions of different SN types to look for implied differences in their parent stellar populations and their progenitor stars.\\
Following \cite{bart92}, \cite{bergh97} found a suggestion that
SNIb/c are more centrally concentrated than SNII, a result that has been confirmed with better statistics by \cite{tsv04} and
most recently \cite{hak08}. This suggests a metallicity dependence in producing SNIb/c, consistent with the scenario where
higher metallicity progenitors have stronger line driven winds (see e.g. \citealt{pul96}; \citealt{kud00}; \citealt{mok07})
and thus lose more of their pre-SN envelopes 
producing a higher fraction of SNIb/c. These studies generally normalise the SN galactocentric distances to the total size
of host galaxies in order to make statistical arguments on differences between the radial distributions of SN types. However,
this analysis does not take into account how the \textit{stellar populations} of these host galaxies are radially distributed 
and therefore complicates the interpretations that can be made on the nature of SN progenitors.\\
Here we use $R$-(or \textit{r'})-band and \ha\ imaging to investigate how the different SN types are radially distributed with respect to
a young stellar population as traced by the \ha\ emission and an older stellar populations as traced by the $R$-(or \textit{r'})-band
continuum light. Using this analysis we address two questions. Firstly, are the radial positions of the different SN types consistent with being drawn at 
random from a young or an old stellar population? 
Secondly, do any of the SN types preferentially occur at particular 
radial positions within their host galaxies and what can this tell us about the nature 
of their progenitor ages and metallicities?\\ 
The paper is arranged as follows: in the next section we present the data and discuss the reduction
techniques employed, in \S ~\ref{rad} we summarise the radial aperture analysis used throughout 
this work, in \S ~\ref{results} we present the results for the radial distribution
of the different SN types, in \S ~\ref{diss} we discuss possible explanations for the results and 
implications that these have on the nature of SN progenitors and finally in \S ~\ref{con} we 
draw our conclusions.

\section{Data}
\label{data}
The initial galaxy sample that formed the data set for JA06 was the \ha\ Galaxy Survey (\ha GS).
This survey was a study of the SF properties of the local Universe using \ha\ imaging of
a representative sample of nearby galaxies, details of which can be found in \cite{jam04}.
63 SNe were found to have occurred in the 327 \ha GS galaxies
through searching the International Astronomical Union (IAU) database
\footnote{http://cfa-www.harvard.edu/iau/lists/Supernovae.html}.\\
Through three observing 
runs on the Isaac Newton Telescope (INT) and a long term time allocation on
the Liverpool Telescope (LT) we have now obtained imaging for the host galaxies of
a total of 178 CC SNe, the analysis of which is presented here. The LT is a fully robotic 2m telescope
operated remotely by Liverpool John Moores University. To obtain our imaging we used 
\textit{RATcam} together with the narrow \ha\ and the broad-band Sloan \textit{r'} filters. Images were binned 
2$\times$2
to give 0.278\as\ size pixels, and the width of the \ha\ filter enabled us to image
target galaxies out to $\sim$2500 \kms. The INT observations used the Wide Field Camera (WFC) together
with the Harris \textit{R}-band filter, plus the rest frame narrow \ha\ (filter 197)
and the redshifted \ha\ (227) filters enabling us to image host galaxies out to $\sim$6000 \kms. During our
2005 INT observing run we also used the SII filter (212) as a redshifted \ha\ filter and imaged 15
SN hosting galaxies at recession velocities of $\sim$7500 \kms. The pixel scale on all INT images is 0.333\as\ per pixel
and with both the LT and INT our exposure times were $\sim$800 sec in \ha\ and $\sim$300 sec in \textit{R} (or \textit{r'},
for the remainder of this paper `$R$-band' will refer to both these filters, and the filter
that was used for each target galaxy is listed in Table~\ref{SNtabrad}1).\\ 
These additional SNe and host galaxies 
were chosen from the Padova-Asiago SN catalogue\footnote{http://web.pd.astro.it/supern/},
as specific SN types were more complete for the listed SNe. 
At a later date all SN type classifications taken from the Padova-Asiago catalogue
were checked through a thorough search of the literature and IAU circulars, as classifications
can often change after the initial discovery and therefore those in the catalogue may not be completely accurate.
The full list of SN types is given in Table~\ref{SNtabrad}1, where references are given if classifications were changed from
those in the above catalogue. The main discrepancies were the classification of the so-called SN `impostors'
as SNIIn in the Padova-Asiago catalogue. These are transient objects that are believed to be the outbursts from very massive Luminous Blue Variable
stars (LBVs), which do not fully destroy the progenitor star and are therefore not classed as true SNe (e.g. \citealt{van00,maun06}).
Four such objects were found in our sample, and the results on these `impostors' are presented and discussed separately
in the following sections.\\
Through the above telescope
time allocations we obtained data on host galaxies of a large fraction of the
discovered CC SNe that have been classified as IIP, IIL, IIb, IIn, Ib, and Ic, and were observable within the \ha\ filters of the two telescopes.\\
There are obvious biases within a set of data chosen in the above way. As we use 
any discovered SNe for our sample, the various different biases in the different SN surveys
that discovered them mean that the galaxy/SN sample is by no means representative of the 
overall SN populations. Bright, well studied galaxies will be over represented, as will brighter
SNe events that are more easily detectable. However, firstly we are not analysing the overall host 
galaxy properties, but
are analysing where within the distribution of stellar populations of the host galaxy the SNe are occurring. 
Secondly, the small number of SNe of each of the CC sub-types that have been discovered means that 
no individual survey can currently manage to analyse the properties of their host galaxies
or parent stellar populations in any statistically significant way 
(most statistical observational studies do not even 
attempt to separate the Ib and Ic SN types). Taking our approach enables us to make statistical
constraints on all the major SN types. The results that are presented 
in this paper are on the analysis of the parent stellar populations of 115 SNII, of which 35 are IIP,
6 IIL, 8 IIb and 12 IIn, 4 SN `impostors', plus 22 Ib, 30 Ic and 6 that only have Ib/c as their classification, taken from
both the initial \ha GS sample and our additional data described above. The results for individual
SNe obtained from the following analysis are listed in Table~\ref{SNtabrad}1, together with host galaxy characteristics.\\

For each SN host galaxy we obtained \ha +\NII\ narrow band imaging, plus \textit{R}-band imaging. 
Standard data reduction (flat-fields, bias subtraction etc) were 
achieved through the automated pipeline of the LT \citep{ste04}, and the 
INT data were processed through the INT Wide Field Camera (WFC), Cambridge Astronomical Survey Unit (CASU)
reduction pipeline. Continuum subtraction
was then achieved by scaling the broad-band image fluxes to those of the \ha\ images using 
stars matched on each image, and subtracting the broad-band images from the \ha\ images. Our reduction
made use of various \textit{Starlink} packages.\\

\section{Radial aperture analysis}
\label{rad}
\begin{figure}
\includegraphics[width=8.5cm]{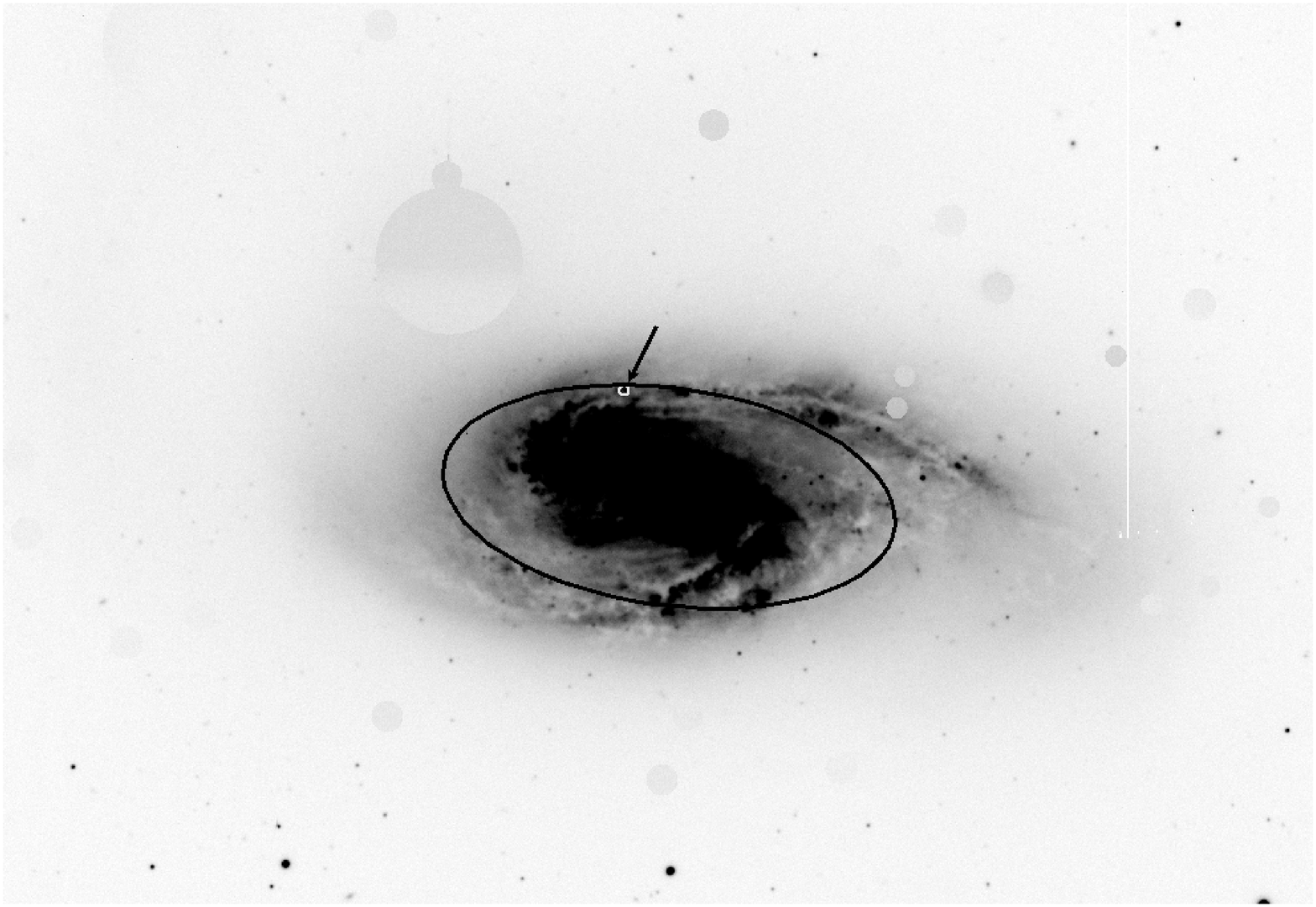}
\includegraphics[width=8.5cm]{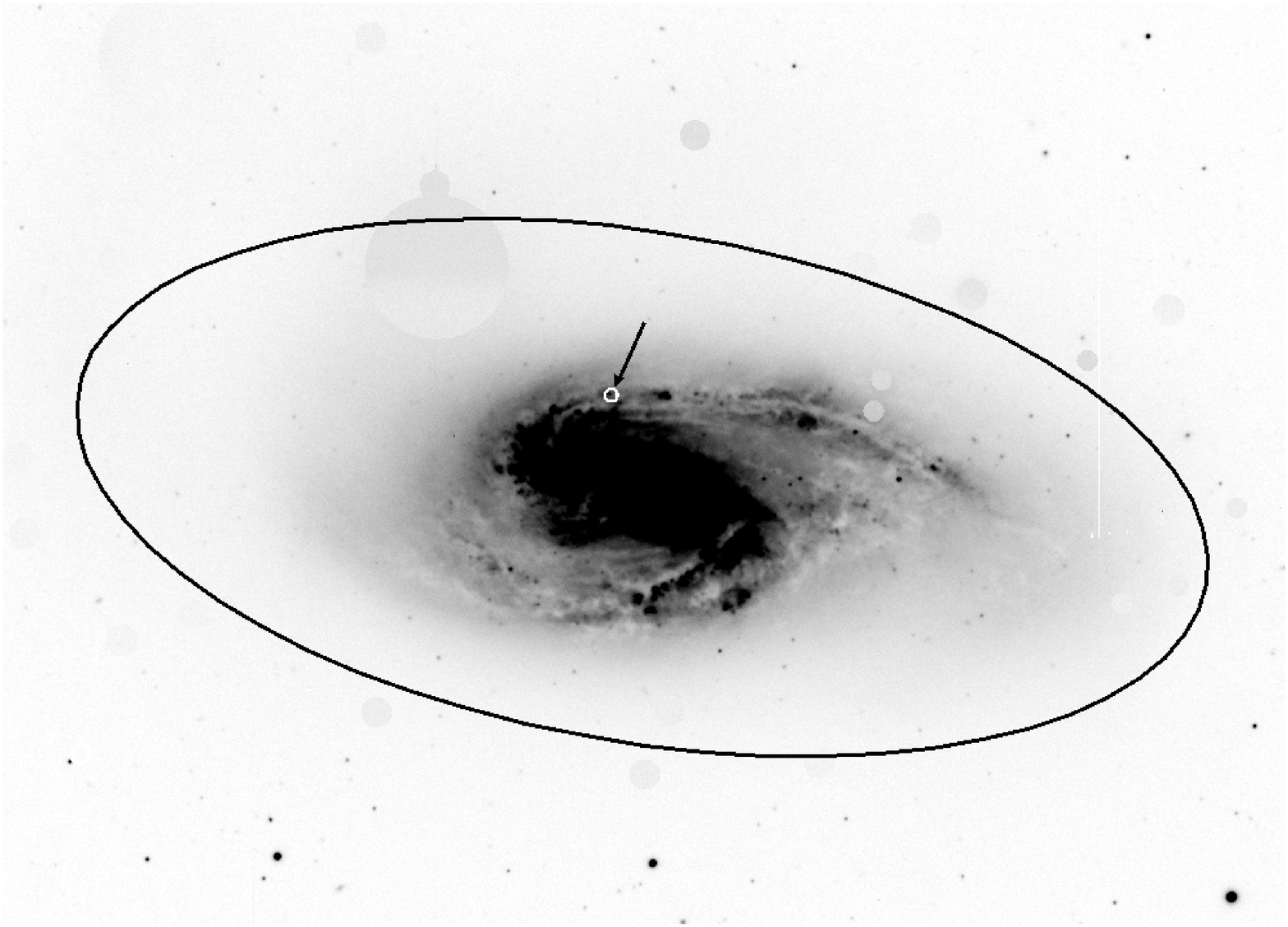}
\caption{An example of the radial aperture analysis applied to NGC 3627, the host galaxy of the IIP SN 1973R. The
top image shows a negative $R$-band image of the galaxy with the position of the SN indicated by the circle and the arrow, and an
aperture overlaid that just includes the SN. The
bottom image shows the same galaxy with an aperture overlaid that contains all the flux of the galaxy. The
fraction of $R$-band light within the position of SN 1973R is 0.471.}
\label{example}
\end{figure}

To analyse where within the radial distribution of $R$-band and \ha\ light SNe occurred, radial growth curves for
both were constructed for the host galaxy images. These were derived in the following way. For both the $R$-band and \ha\
images all apertures were centred on the $R$-band galaxy centroid positions. Galaxy parameters (semi-major and  -minor axis sizes, plus position angles)
were taken from the NASA/IPAC
Extragalactic Database (NED)\footnote{http://nedwww.ipac.caltech.edu/}. 
For face-on spiral and irregular galaxies circular 
apertures were produced, while for inclined spirals elliptical apertures were produced using the ratio of the galaxy axes taken from NED.
These were then checked on each host 
galaxy image by eye to ensure they were a good fit to the sample images. Concentric apertures were then produced 
that increased in semi-major axis by 5\as\ and sampled the galaxy light out to sky levels. 
In general 20 to 100 apertures were 
required to sample the full extent of the galaxy light. For those galaxies where less than 25 apertures were required (43 host galaxies) 
concentric apertures that increased by 2.5\as\ in semi-major axis were used so that a significant sampling of radial positions was
achieved.
These concentric apertures were then used to produce radial growth curves for each SN host galaxy by
calculating the flux within each ellipse of increasing size. This enables one to define the edge of the host galaxy at the point 
where the curves flatten off as the aperture counts have reached sky levels.\\
The next step was to calculate the total flux within the aperture just
including the SN position.
This aperture was calculated using the SN coordinates relative to the galaxy centre, plus both
the position and inclination angles of the host galaxy (the value calculated is therefore the
semi-major axis of the ellipse that just includes the SN position, taking into account the projection along 
the minor axis). 
Using these apertures one can then calculate the fraction of the total galaxy $R$-band light and \ha\ emission (\textit{Fr}$_{\textit{R}}$ and \textit{Fr}$_{\textit{\ha}}$
as they will be referred to in the results) that is within the position of each SN.
This is achieved by measuring the flux within the SN-containing ellipse and dividing this by the measured flux within the
largest ellipse (i.e. the total galaxy flux).
In Figure~\ref{example} we show an example of this technique at work, where two $R$-band images
of the SN host galaxy NGC 3627 are presented. The top image shows the position of the SNIIP 1973R and the aperture that just includes the
SN, while the bottom image has overlaid the aperture that contains all of the $R$-band flux of the galaxy.\\
During this analysis it was found that there were some SNe where it was not possible to perform the 
above procedure accurately. Some of the $R$-band images were not used due to there being 
too many bright stars in the field as it can be difficult to 
remove stars from images without leaving residuals that may affect the results. This in general is not a 
problem for the \ha\ images as stars are removed through the continuum subtraction applied. There
were also images where due to the size of the field of view of the detector used (mainly LT images), it was not 
possible to apply the radial analysis out the full size of the host
galaxies and therefore these data were removed from the subsequent
analysis and results.\\

The method described above normalises the distributions to the host galaxy properties (overall
$R$-band light and \ha\ emission fluxes). This enables an analysis of a sample of SNe that occur in a wide range
of galaxy morphologies and sizes, without the characteristics of those host galaxies affecting the overall distribution of
SN positions.
If one were to merely use absolute SN distance from its host galaxy centre (in say, kpc)
then there would be a systematic overabundance of SNe occurring at small radial positions as these
positions would be populated with SNe exploding in all host galaxies, while larger radial positions
would only be populated by SNe occurring in the largest host galaxies. This
normalisation also allows an investigation into whether the different SN types trace the 
overall distributions of the two types of light within their host galaxies, as discussed in \S ~\ref{SNrad}.\\ 
If the SNe are equally likely to arise from any part of the \ha\ emission or $R$-band light distributions then one would expect a mean fraction of 0.5 
and the fractions of light to be evenly distributed throughout the radial light profiles (i.e. a flat distribution). This is the test
that is presented in the following section; whether the different SN types better trace the 
radial distribution of the $R$-band light or the \ha\ emission. It is also investigated whether there 
are differences between the radial distributions of the different SN types and what
implications these may have for the nature of their progenitors.

\begin{figure}\centering
\includegraphics[width=8cm]{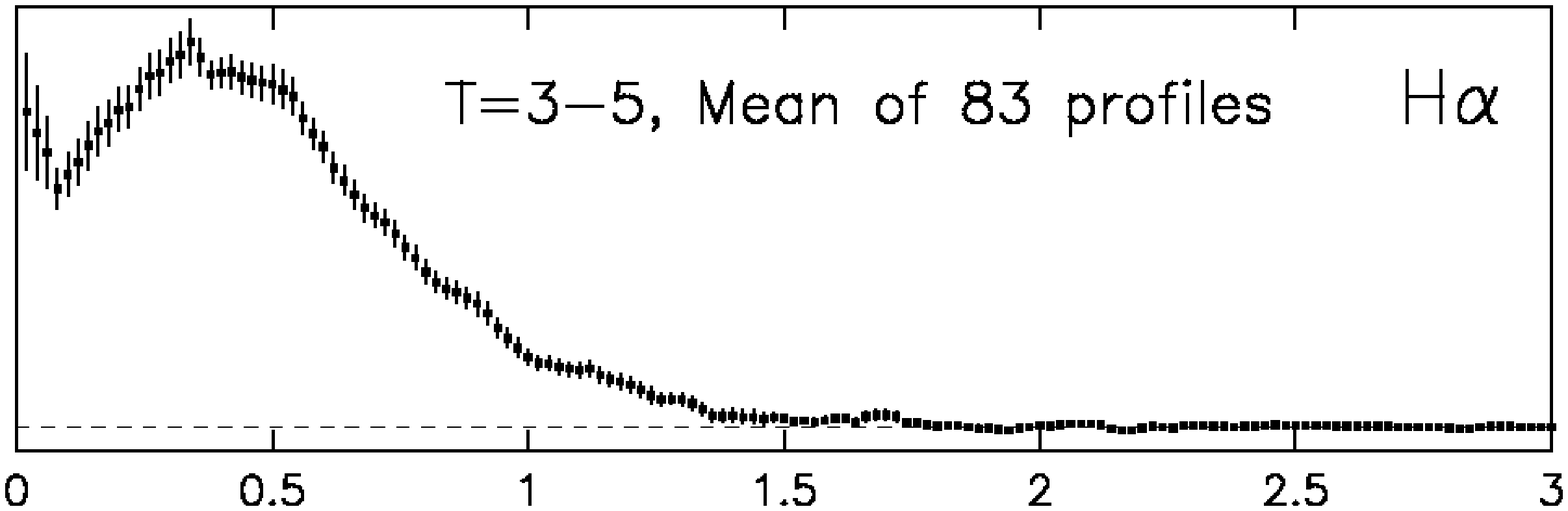}
\includegraphics[width=8cm]{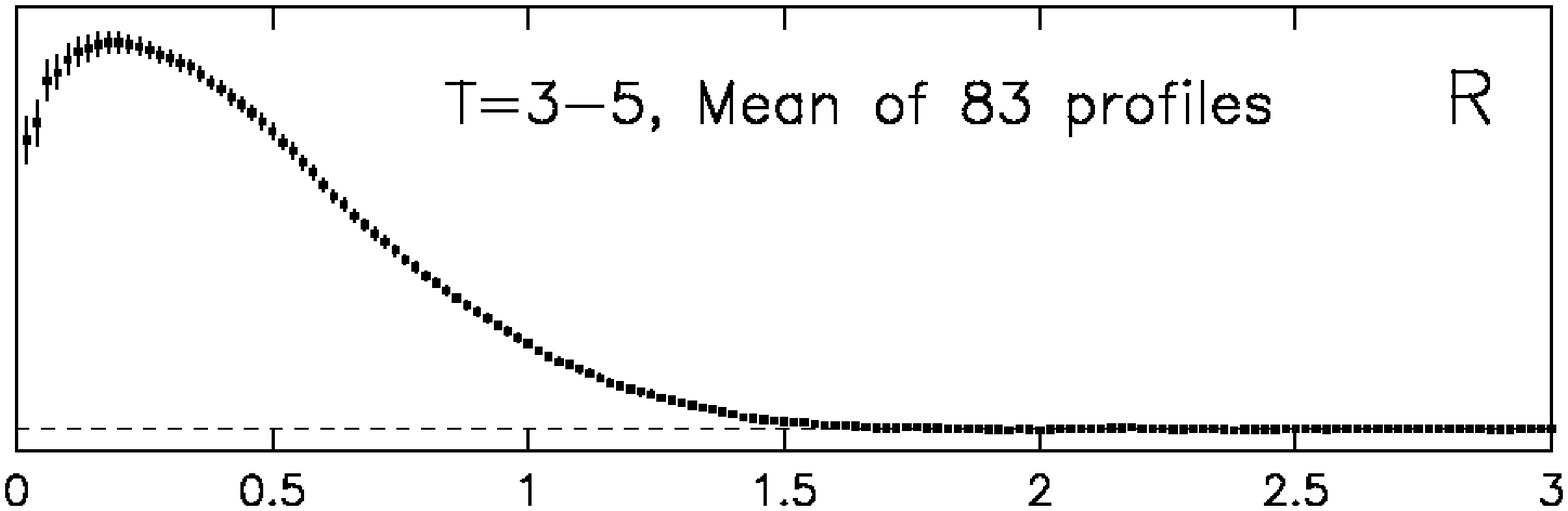}
\includegraphics[width=8cm]{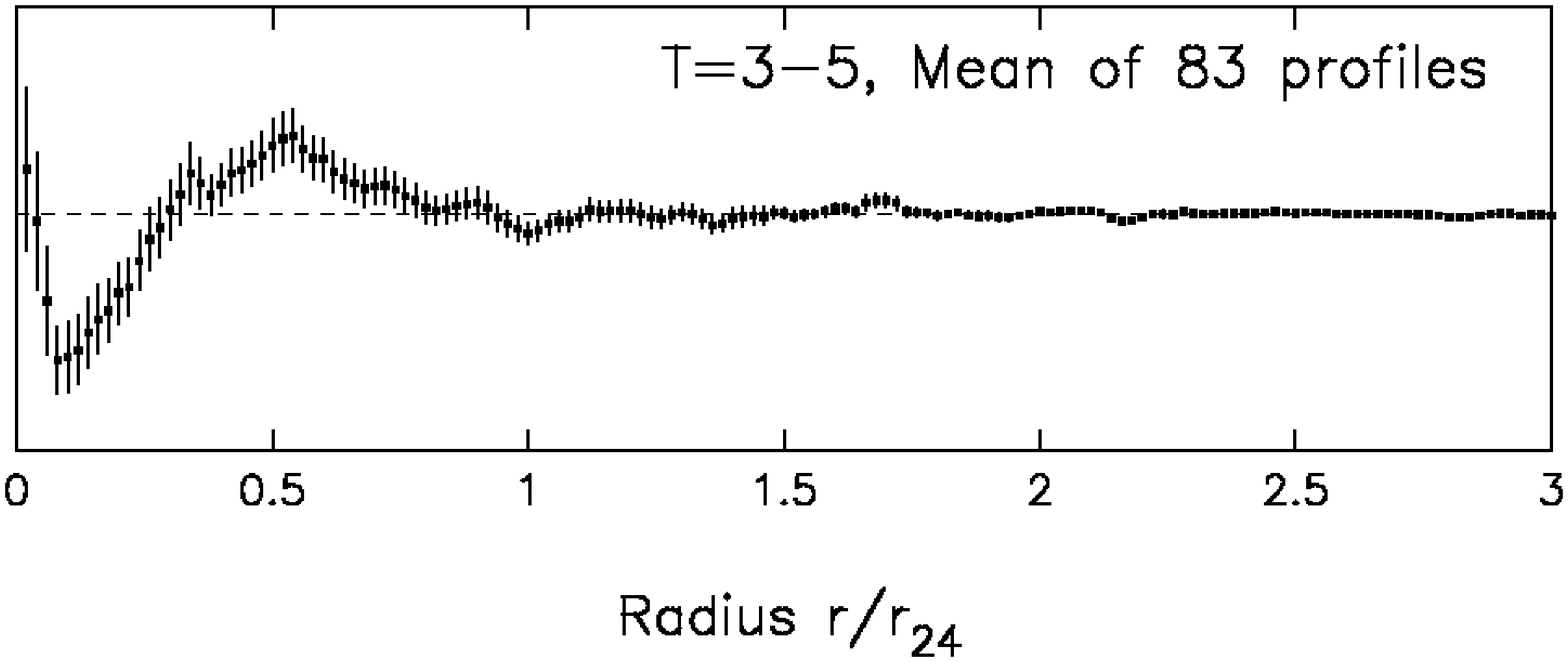}
\caption{Normalised light profiles for all the \ha GS galaxies of types Sb, Sbc, and Sc (83 in total, figures modified from those
in \citealt{jam09}). The top panel shows the radial \ha\ profile averaged over the sample and the middle panel shows the $R$-band profile.
In the bottom panel we show the difference between the mean \ha\ and $R$-band profiles.}
\label{radfigs}
\end{figure}

In Fig.~\ref{radfigs} we show the normalised radial light profiles (flux in elliptical annuli), averaged over all
of the T-types 3-5 galaxies (Sb, Sbc and Sc Hubble types, both barred and unbarred)
within the \ha GS survey. This range of types
is chosen to illustrate the typical light profiles for galaxies used in the present investigation, as they represent
those galaxies that dominate the SF within the local Universe \citep{jame08}, and also dominate the 
morphological types within our sample (as shown in \S ~\ref{galclass}). In the top panel we show the 
averaged  \ha\ profile, in the middle panel we show the averaged $R$-band profile, and finally in the
bottom panel we show the difference between the two. In the final plot the central spike is possibly due to central SF or
AGN in barred galaxies, while the negative values between r/r$_{24}$ of 0.1 and 0.3 show the effect 
of the central bulge, where there is a deficit in the SF relative to the $R$-band light. For a detailed discussion of the calculation of these profiles, and in particular the
scaling of the input profiles by radius and total flux, see \cite{jam09}.

\section{Results}
\label{results}

\subsection{SNII}
\label{II}
\begin{figure}\centering
\includegraphics[width=8cm]{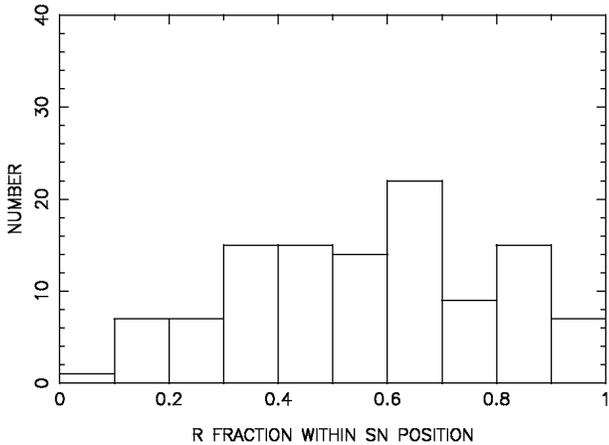}
\caption{Histogram of the fractional $R$-band fluxes within the locations of 113 SNII}
\label{figradIIR}
\end{figure}

\begin{figure}\centering
\includegraphics[width=8cm]{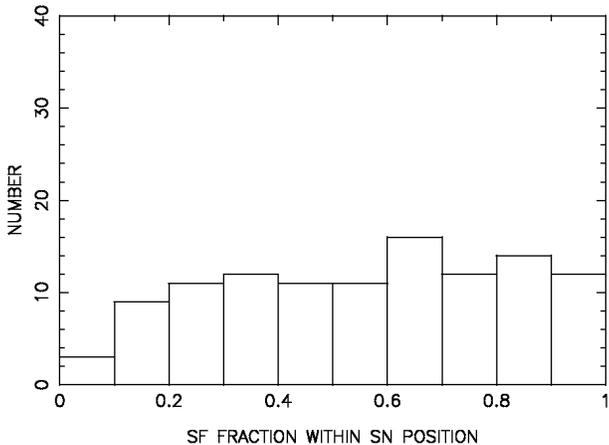}
\caption{Histogram of the fractional \ha\ fluxes within the locations of 113 SNII}
\label{figradIIHa}
\end{figure}

The mean fraction of $R$-band light (\textit{Fr}$_{\textit{R}}$ henceforth)
within the positions of the 113 SNII is 0.566 (standard error on the mean of 0.022) while for the \ha\ emission 
(\textit{Fr}$_{\textit{\ha}}$ henceforth) it is 0.572 (0.025). Histograms of these two distributions are shown in 
Figs.~\ref{figradIIR} and~\ref{figradIIHa} respectively. Using a Kolmogorov Smirnov (KS) test there is only $\sim$0.1\%\ chance that
the SNII are drawn from the same radial distribution as the $R$-band light and
there seems to be a deficit at both small and large radii. Although the distribution
of SNII compared to the \ha\ emission appears reasonably flat in Fig.~\ref{figradIIHa}, there is only a 5\%\ chance 
that they follow radial distribution of the HII regions. This is due to the
deficit of SNe at small radii which will be discussed in the next section. Results on 
the individual SNII sub-types will now be presented.

\subsubsection{SNIIP}
\label{IIP}
\begin{figure}\centering
\includegraphics[width=8cm]{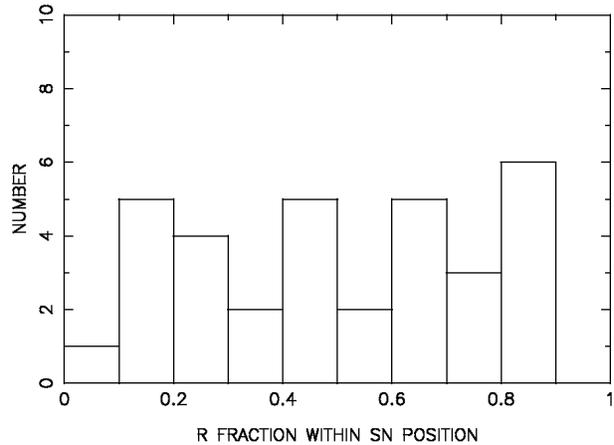}
\caption{Histogram of the fractional $R$-band fluxes within the locations of 34 SNIIP}
\label{figradIIPR}
\end{figure}

\begin{figure}\centering
\includegraphics[width=8cm]{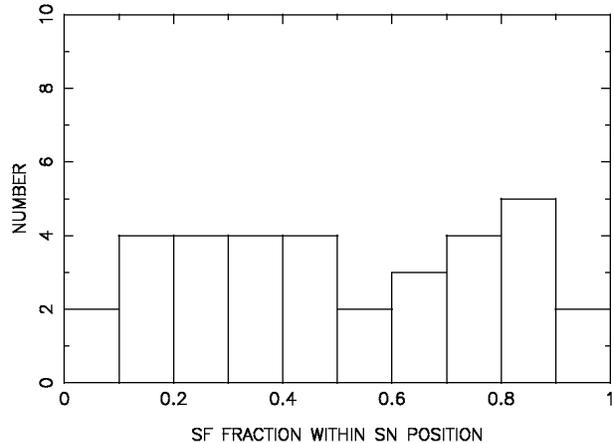}
\caption{Histogram of the fractional \ha\ fluxes within the locations of 35 SNIIP}
\label{figradIIPHa}
\end{figure}

The \textit{Fr}$_{\textit{R}}$ and \textit{Fr}$_{\textit{\ha}}$ distributions for the SNIIP are shown in Figs.~\ref{figradIIPR} and ~\ref{figradIIPHa}
respectively. Both of these distributions appear similar and are consistent with the SNIIP being 
drawn from the same population as both the $R$-band and \ha\ emission  (using a KS test),
although there appears to be slightly more of a deficit at small and large radii
for the SNe with respect to the $R$-band light, as seen for the overall SNII population. The mean \textit{Fr}$_{\textit{R}}$ is
0.511 (0.046) and the mean \textit{Fr}$_{\textit{\ha}}$ is 0.522 (0.050). Both of these distributions are also consistent with being 
drawn from the same population as the overall SNII values.

\subsubsection{Other SNII sub-types}
\label{subtypes}
In Table 1 we list the radial statistics for
the SN types IIL, IIb, IIn and `impostors'. All these distributions
are consistent with being drawn from the same parent population as the overall
SNII. However, we note that all distributions are shifted to higher mean values
%located slightly further
%away from the centres of the light distributions of their hosts 
than the SNII. 
The `impostors' 
in particular seem to be located towards the outer parts of their host galaxies.

\begin{table}
\label{subs} 
\centering
\begin{tabular}[t]{cccccc}
\hline
\hline
SN type & N & $\bar{F}$$r$$_{\textit{R}}$& $\sigma$$_M$ & $\bar{F}$$r$$_{\textit{\ha}}$ & $\sigma$$_M$  \\
\hline
IIL & 6 & 0.520 & 0.074 & 0.561 & 0.101\\
IIb & 8 & 0.581 & 0.106 & 0.577 & 0.116\\    	     
IIn & 12 & 0.591 & 0.060 & 0.580 & 0.089\\
`impostors' & 4 & 0.725 & 0.150 & 0.757 & 0.168\\ 
\hline
\hline
\end{tabular}
\caption{Radial statistics for the SN sub-types; types IIL, IIb, IIn and
`impostors'. In the first two columns the SN type and number of events in the sample
are given. In columns 3 and 4 
the mean \textit{Fr}$_{\textit{R}}$ and its associated error are listed followed by the 
statistics for the \textit{Fr}$_{\textit{\ha}}$.}		
\end{table}

%\subsubsection{SNIIL}
%\label{IIL}
%The mean \textit{Fr}$_{\textit{R}}$ for the 6 SNIIL is 0.520 (0.074) and the mean 
%\textit{Fr}$_{\textit{\ha}}$ is 0.561 (0.101). These SNe appear to be drawn from
%the same distributions as the overall SNII.

%\subsubsection{SNIIb}
%\label{IIb}
%The mean \textit{Fr}$_{\textit{R}}$ for the 8 SNIIb is 0.581 (0.106) and the mean 
%\textit{Fr}$_{\textit{\ha}}$ (only 7) is 0.577 (0.116), again consistent with 
%the distributions of the overall SNII.
%
%\subsubsection{SNIIn}
%\label{IIn}
%The mean \textit{Fr}$_{\textit{R}}$ for the 12 SNIIn is 0.591 (0.060) and the mean 
%\textit{Fr}$_{\textit{\ha}}$ is 0.580 (0.089), again consistent with 
%the distributions of the overall SNII.
%
%\subsubsection{SN `impostors'}
%\label{imp}
%The mean \textit{Fr}$_{\textit{R}}$ for the 4 SN `impostors' is 0.725 (0.150) and the mean 
%\textit{Fr}$_{\textit{\ha}}$ is 0.757 (0.168). Therefore these events seem to result from 
%progenitors located towards the outer parts of their host galaxies.

\subsubsection{Comparison of SNIIP to other sub-types}
\label{compareIIP}
\begin{figure}\centering
\includegraphics[width=8cm]{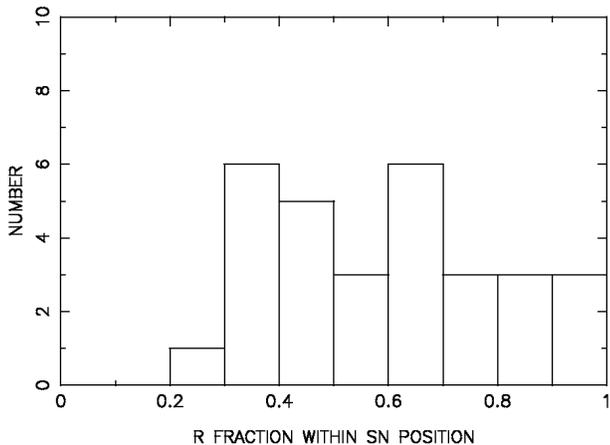}
\caption{Histogram of the fractional $R$-band fluxes within the locations of the combined
sample of SNIIL, IIb, IIn and SN `impostors' (30 objects in total).}
\label{nonPR}
\end{figure}

\begin{figure}\centering
\includegraphics[width=8cm]{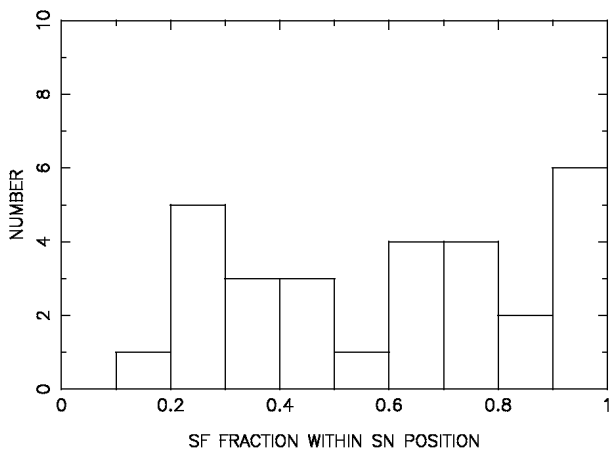}
\caption{Histogram of the fractional \ha\ fluxes within the locations of the combined
sample of SNIIL, IIb, IIn and SN `impostors' (29 objects in total).}
\label{nonPHa}
\end{figure}

For a comparison of the SNII sub-types with the individual SNIIP presented above,
in Figs.~\ref{nonPR} and~\ref{nonPHa} the distributions of the combined sample of the 
SNIIL, IIb and the IIn plus the four SN `impostors' are presented. It is
immediately clear that there is a large deficit of these non-IIP SNe at small
radii compared to the SNIIP (see Figs.~\ref{figradIIPR} and~\ref{figradIIPHa}), an effect that seems more prominent
in the $R$-band. However, using a KS test both of the distributions are consistent with those of the SNIIP. 
The non-IIP SNe have a $\sim$3\%\ chance of being drawn from the distribution of $R$-band light (i.e. a flat
distribution) while they are consistent with being drawn from the \ha\ emission distribution. The mean 
\textit{Fr}$_{\textit{R}}$ for the 30 non-IIP SNe is 0.592 (0.040) and the mean 
\textit{Fr}$_{\textit{\ha}}$ (only 29) is 0.600 (0.050)

\subsection{SNIb/c}
\label{Ibc}
\begin{figure}\centering
\includegraphics[width=8cm]{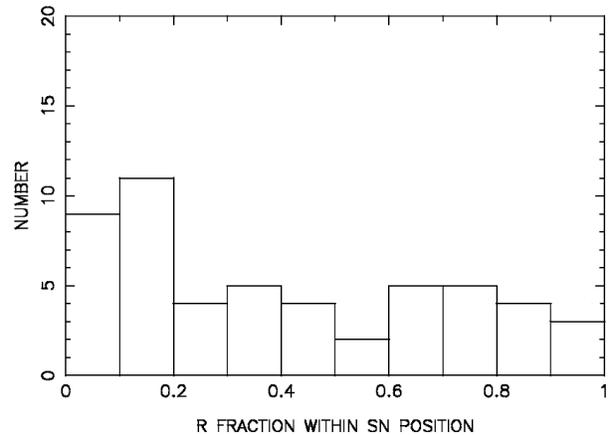}
\caption{Histogram of the fractional $R$-band fluxes within the locations of 52 SNIb/c}
\label{figradIbcR}
\end{figure}

\begin{figure}\centering
\includegraphics[width=8cm]{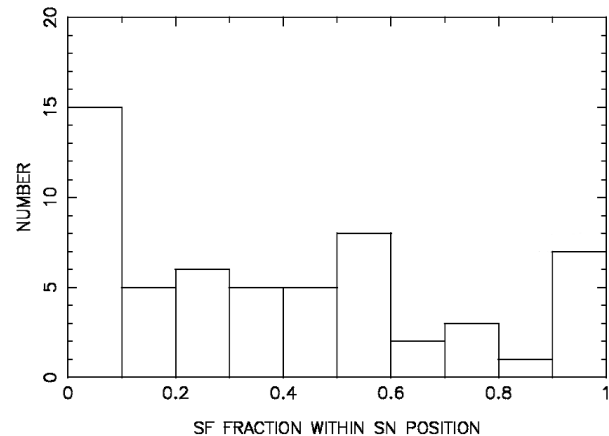}
\caption{Histogram of the fractional \ha\ fluxes within the locations of 58 SNIb/c}
\label{figradIbcHa}
\end{figure}

The radial distributions of the overall SNIb/c population with respect to the $R$-band light and \ha\ emission are shown in 
Figs.~\ref{figradIbcR} and~\ref{figradIbcHa} respectively (52 SNe in the $R$-band histogram and 58 in the \ha\ plot). It is
immediately clear that they are weighted to the central parts of the light distribution, especially when comparing them to the
other SN types discussed above. This is also reflected in the mean values of their distributions; the mean \textit{Fr}$_{\textit{R}}$ is
0.408 (0.041) and mean \textit{Fr}$_{\textit{\ha}}$ is 0.405 (0.041). Comparing these distributions to that of the 
overall SNII population it is found that for both the $R$-band and the \ha\ there is only $\sim$0.1\%\ chance that
the SNIb/c and SNII are drawn from the same radial distributions. Comparing the above samples to flat distributions it 
is found that there is only $\sim$2\%\
chance that the SNIb/c are drawn from the same radial distribution as that of the SF while there is only $\sim$1\%\ chance that they are drawn from the 
radial distribution of the $R$-band light. Results on the SNIb and Ic will now be presented separately.

\subsubsection{SNIb}
\label{Ib}
\begin{figure}\centering
\includegraphics[width=8cm]{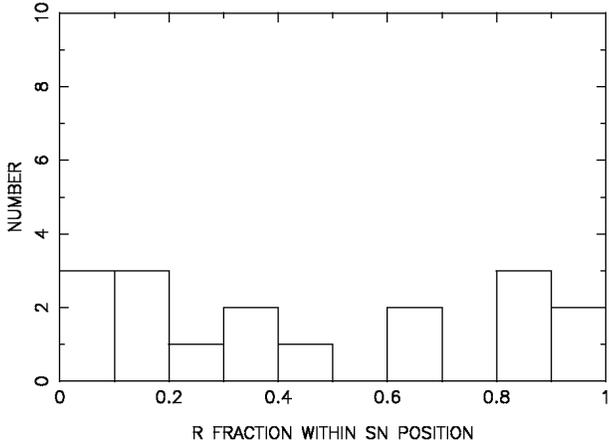}
\caption{Histogram of the fractional $R$-band fluxes within the locations of 17 SNIb}
\label{figradIbR}
\end{figure}

\begin{figure}\centering
\includegraphics[width=8cm]{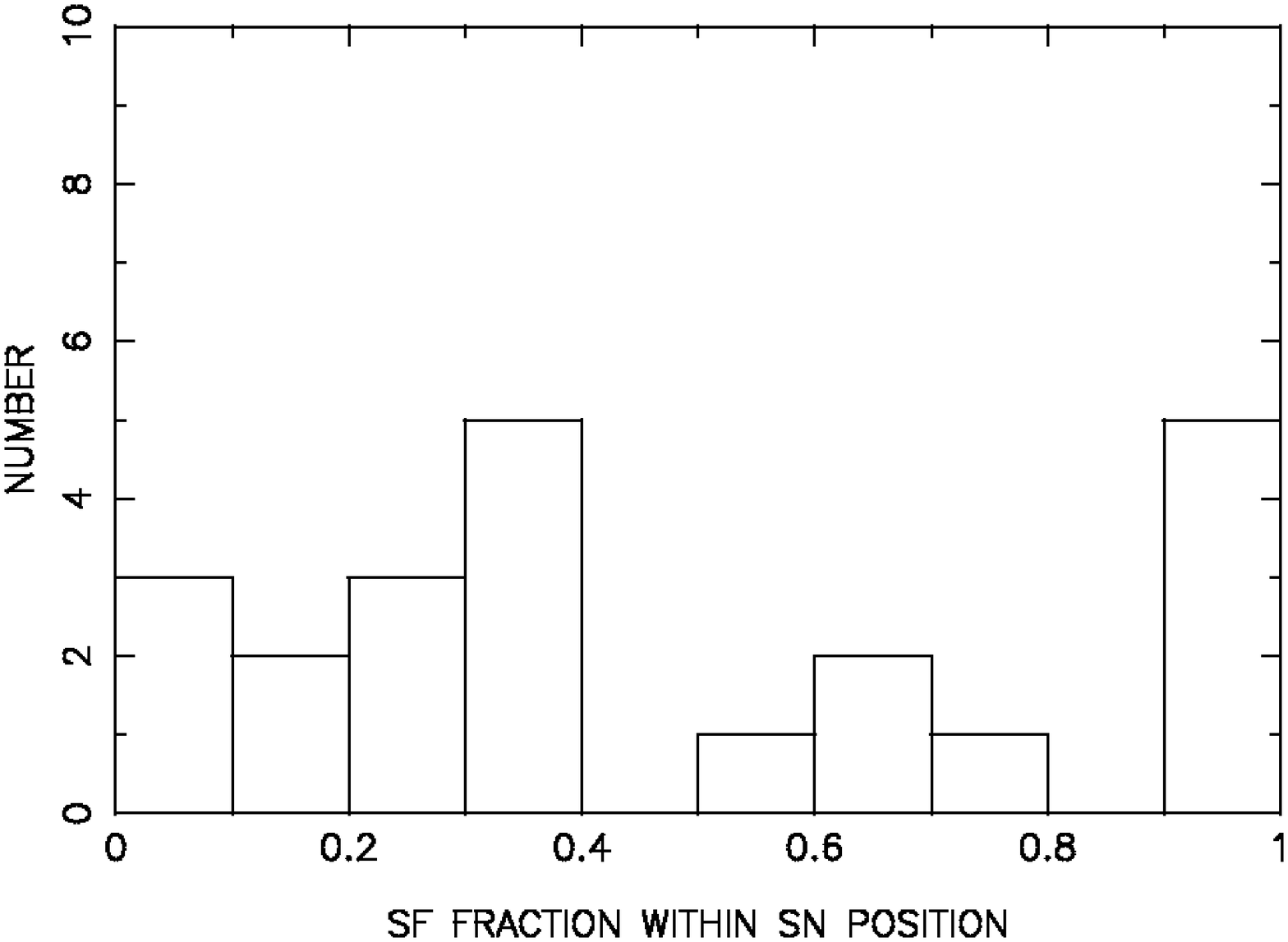}
\caption{Histogram of the fractional \ha\ fluxes within the locations of 22 SNIb}
\label{figradIbHa}
\end{figure}

The mean \textit{Fr}$_{\textit{R}}$ for the 17 SNIb is 0.468 (0.088) and the mean 
\textit{Fr}$_{\textit{\ha}}$ (22 SNe) is 0.465 (0.072) and histograms
of the two distributions are presented in Figs.~\ref{figradIbR} and~\ref{figradIbHa} respectively. These SNe
are consistent with being drawn from the same radial distribution as that of the $R$-band light and the 
\ha\ emission of their host galaxies. There is only a $\sim$2\%\ ($R$-band) and $\sim$7\%\ (\ha ) chance
that the SNIb are drawn from the same radial distributions as the SNII.

\begin{figure}\centering
\includegraphics[width=8cm]{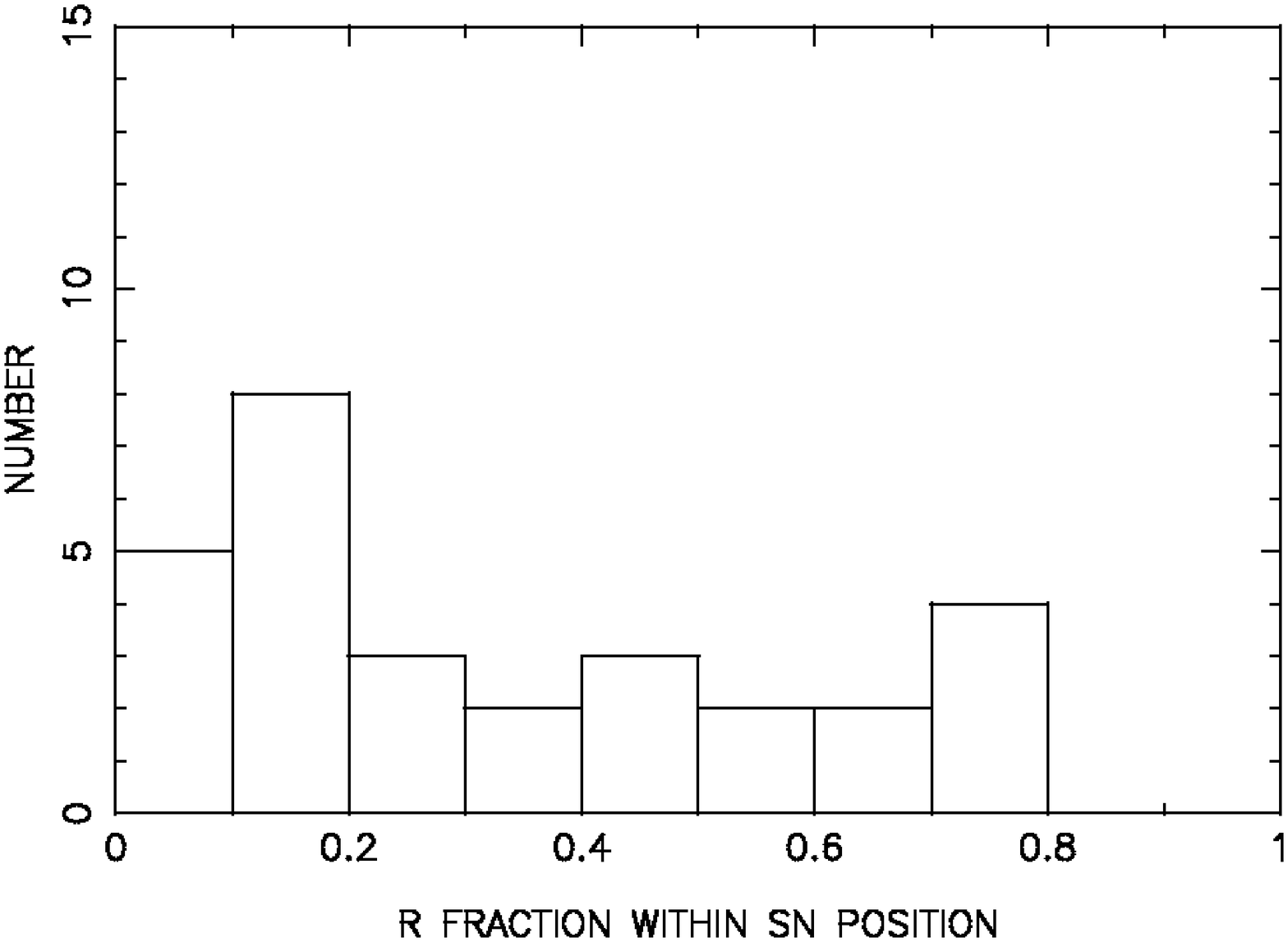}
\caption{Histogram of the fractional $R$-band fluxes within the locations of 29 SNIc}
\label{figradIcR}
\end{figure}
\begin{figure}\centering
\includegraphics[width=8cm]{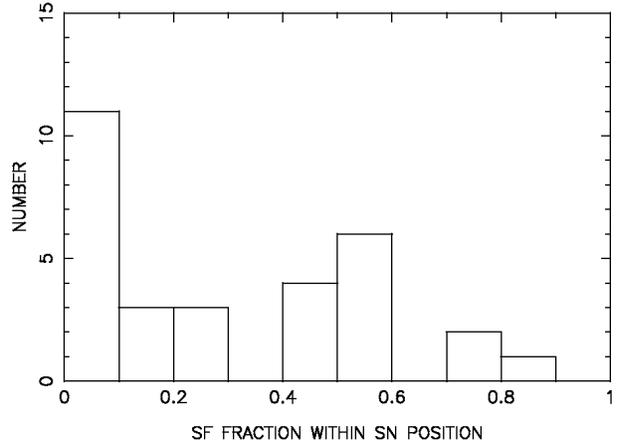}
\caption{Histogram of the fractional \ha\ fluxes within the locations of 30 SNIc}
\label{figradIcHa}
\end{figure}

\subsubsection{SNIc}
\label{Ic}
The \textit{Fr}$_{\textit{R}}$ (29 SNe) and \textit{Fr}$_{\textit{\ha}}$ (30) distributions for the SNIc are shown in Figs.~\ref{figradIcR} and ~\ref{figradIcHa}
respectively.
The mean \textit{Fr}$_{\textit{R}}$  is 0.338 (0.047) and the mean 
\textit{Fr}$_{\textit{\ha}}$ is 0.306 (0.047). The histograms show that both these distributions are strongly weighted to
small light fractions; i.e. the central parts of the light distributions of host galaxies. This is are backed up by KS tests that show there 
is only $\sim$1\%\ chance that the SNIc follow the same radial distributions as that of the $R$-band light, while there is
$<$1\%\ chance that they follow the same radial distribution as the line emission. Comparing these distributions 
to those of the SNIb it is found that while when looking at their distributions (specifically comparing Figs.~\ref{figradIbHa}
and~\ref{figradIcHa}, and the mean \textit{Fr}$_{\textit{\ha}}$ values) they seem to occur at different host galaxy radii (particularly
with respect to the complete deficit of SNIc in the 
outer 20\%\ of the $R$ band light distribution), statistically they are consistent with being drawn 
from the same parent distribution.

\section{Discussion}
\label{diss}
\subsection{Implications for progenitor metallicity}
\label{Z}
As discussed in \S ~\ref{SNrad}, radial position within galaxies can be used as a proxy for
environment metallicity with galaxy centres tending to have higher metal content. Differences in the radial
distributions of SN types can therefore be used to infer differences in progenitor metallicity. Comparing Fig.~\ref{figradIbcR}
to Fig.~\ref{figradIIR} it is immediately clear that the SNIb/c are more centrally concentrated than the SNII 
(a result also seen in \citealt{bergh97}, and most recently \citealt{hak08})
implying
higher metallicity progenitors.
This trend is also backed up by the KS statistics presented above. 
\begin{figure*}\centering
\includegraphics[width=13cm]{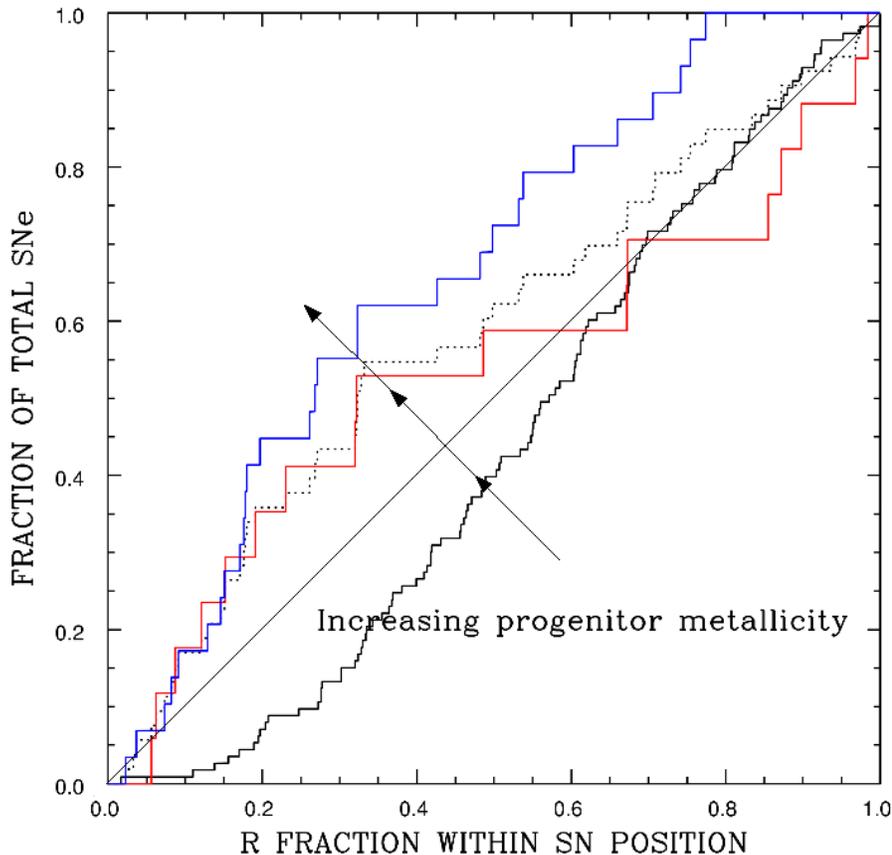}
\caption{Plot showing the cumulative distributions of the radial positions of the main CC SN types with respect
to the $R$-band light of their host galaxies. The SNII distribution is plotted in black,
the SNIb in red, the SNIc distribution is blue and the overall SNIb/c population in dotted black. There is also a hypothetical 
flat distribution (i.e. one that accurately traces the radial distribution of $R$-band light) plotted as a diagonal black
line.}
\label{radmetR}
\end{figure*}
To further illustrate this point, in Fig.~\ref{radmetR}
the cumulative distributions of the radial positions of the main CC SN types (II, Ib and Ic) with respect to the
$R$-band light are plotted (the $R$-band light distribution is chosen to show this result as it will be a more 
reliable indicator of integrated SF over the galaxy's lifetime, and therefore metallicity, than the distribution of
\ha\ emission). Apart from a central deficit of SNII these SNe seem to follow the distribution of $R$-band light while the SNIb
then the SNIc appear more centrally concentrated as their distributions are biased to smaller radii. This 
plot implies that the SNIb/c tend to preferentially occur in
regions of higher metallicity than SNII. The arrow on the plot shows the implied direction 
of increasing metallicity and one can see that a trend is observed going
from SNII to SNIb and finally SNIc arising from progenitors of the highest metallicity. 
Stars 
of higher metallicity tend to have stronger radiatively driven stellar winds (e.g. \citealt{pul96}; \citealt{kud00}; \citealt{mok07}). 
As discussed in \S ~\ref{SNprog} the spectral appearance of SNIb suggests that these SNe have lost their 
hydrogen envelopes while SNIc spectra indicate that they have also lost their helium envelopes. The results presented above are therefore
consistent with progenitor metallicity playing a significant role in removing SN progenitor envelopes 
and producing the different CC SNe that we see.\\
This metallicity dependence has also been seen in research looking at the host galaxy
properties of CC SNe. \cite{pran03} looked at the relative rates of
SNII and SNIb/c as a function of host galaxy luminosity, using this as
a proxy for metallicity. They found that the ratio of SNII to SNIb/c
events strongly correlated with host luminosity in the sense
that relatively more SNIb/c were found in more luminous, i.e. more
metal rich environments. \cite{pri08_2} investigated the
metallicity of SDSS galaxies that had hosted CC SNe using spectra and came to similar
conclusions. While our results are consistent with these findings, they
would also suggest that the metallicity implications that these authors discuss
are underestimates of the true differences between progenitor
metallicity of SNIb/c and SNII (as \citealt{pri08_2} point out). This is due to the fact that while SNIb/c seem to arise from
higher metallicity host galaxies, they are also more centrally concentrated,
and therefore within these host galaxies are likely to come from higher metallicity stellar populations than SNII as well.
These
different studies (including the current analysis) are also
consistent with the model predictions of single stellar models
(e.g. \citealt{heg03}, \citealt{eld04}), 
that SNIb/c generally arise from more metal rich progenitors. One
point that we emphasise here is that these studies (and others), plus stellar models
tend to group SNIb and SNIc together when discussing
differences in progenitor characteristics. We find that when splitting
these SNe into their Ib and Ic classifications there appears to be a
trend in progenitor characteristics that we observe, implied from the results presented both
here and in AJ08.
This trend implies that SNIb tend to arise from higher mass and more metal rich
progenitors than SNII, while SNIc arise from still higher mass and
metallicity progenitors.\\
{We note that while the statistics that we present here, plus
those from elsewhere suggest that \textit{on average} SNIc
arise from metal rich stellar populations, a number of cases that do not fit this trend have
been observed. Examples of these can be found in \cite{mod08}
and \cite{pri08_2}. Therefore while
SNIc seem to favour high metallicity environments,
it seems that these events can arise from low metallicity progenitors.

\subsection{Comparison of SN and stellar radial distributions}
\label{radpop}
Although the overall SNII population follows the distribution of SF as indicated in Fig.~\ref{figradIIHa} (as would be expected for SNe with massive progenitors),
there is a substantial deficit at small radii, a result that was also seen in JA06. 
One may initially presume that this is due to the bulge 
components of galaxies within their centres that are too old to produce SNII. However, this distribution is \textit{with respect
to the SF of the host galaxies} and therefore there \textit{is} a fraction of high mass SF (the central tenth of the emission) that
is not producing its share of SNII. One possible explanation is the decreasing efficiency of detecting SNe in the central
parts of galaxies due to either the high surface brightness of the stellar background (\citealt{shaw79}) or
extinction effects which have been deduced to result in significant deficits in SN detection rates, even in 
near-IR searches (\citealt{mann03}). 
Another
explanation is that there is an additional emission-line component at the centres of galaxies that is not 
related to SF and hence not linked to SNII. However, both of these explanations would then seem in contradiction
to the above findings that SNIb/c are \textit{more} centrally concentrated than would be expected if they accurately traced
the emission, which is also seen with respect to
the $R$-band light (see Figs.~\ref{figradIbcR} and~\ref{figradIbcHa}), as there seems
no reason why these effects would apply to the SNII but not the SNIb/c (although the possible effects of differences
in intrinsic SN brightness will be discussed below). If one then follows the above
arguments on radial positions being indicators of metallicity of stellar environment, this 
then implies that there is some metallicity upper limit, above which SNII tend not to occur in significant
numbers. The stellar models of \cite{eld04} predict that the minimum initial mass for producing SNII
increases with metallicity. Therefore the increased metallicity in galaxy centres may be 
expected to produce a smaller fraction of SNII than in the lower metallicity outer disk regions (although note that the models
of \citealt{heg03}, do not predict a change in the minimum initial mass). 
Therefore the most plausible explanation for the above result
seems to be that there is some physical mechanism that favours the production of SNIb/c and also 
limits the production of SNII in the centres of galaxies.\\
A remarkable feature of Fig.~\ref{figradIcR} is that 13 out of the 29 SNIc arise from the central 20\%\
of the $R$-band luminosity of these galaxies, which would generally be assumed to be dominated
by the passive, non-star forming bulge population.\\

\subsection{The overall CC SN distribution and correlations with host
  galaxy star formation}
\label{CCdiss}

%\label{CCoverall}
%\begin{figure}\centering
%\includegraphics[width=8cm]{CC_IIR_166.eps}
%\caption{Histogram of the fractional $R$-band fluxes within the locations of the overall CC SN population. The overall
%CC population are plotted in the solid histogram while the SNII are plotted as the dashed histogram.}
%\label{figCCR}
%\end{figure}

\begin{figure*}\centering
\includegraphics[width=13cm]{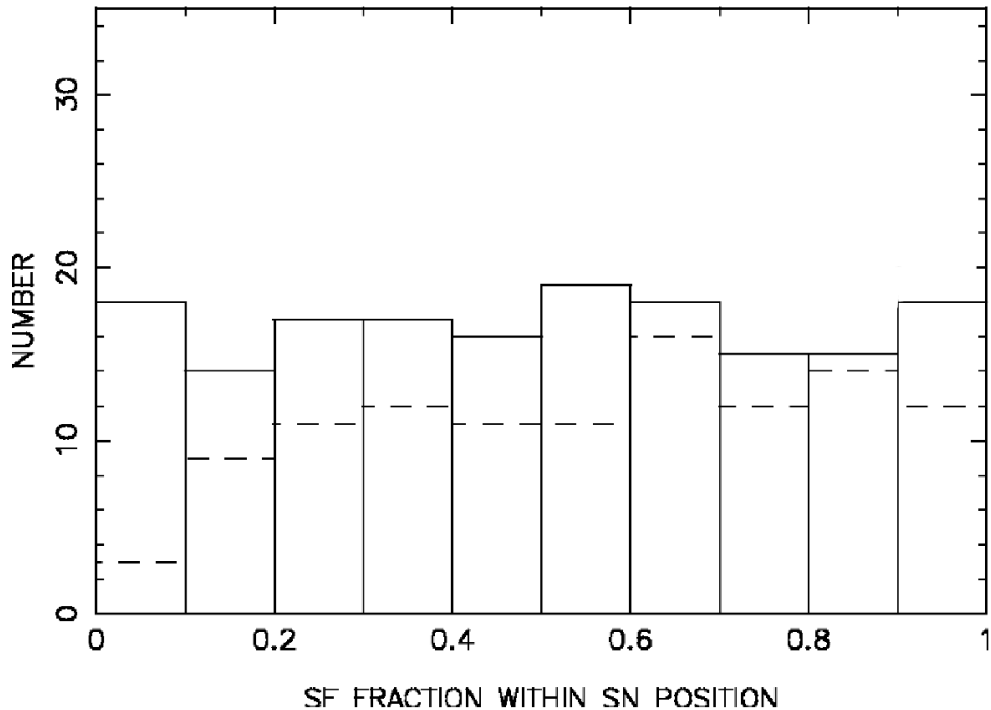}
\caption{Histogram of the fractional \ha\ fluxes within the locations of the overall CC SN population. The overall
CC population are plotted in the solid histogram while the SNII are plotted as the dashed histogram.}
\label{figCCHa}
\end{figure*}

%The above results and discussion suggest that there is a deficit in the number of SNII in the central parts
%of both the radial distribution of the $R$-band light and the \ha\ emission. At the same time there is also
%an excess in the number of SNIb/c seen at the same small radii. 
%The overall CC SN population are believed to arise
%from massive stars with initial masses $>$8\msun\ therefore one would expect their SNe to follow the same radial distribution as
%that of the \ha\ emission, which is thought to arise from similar mass stars. 
In Fig.~\ref{figCCHa} we show the distributions of locations of all types of CC SNe combined, together
with the individual SNII distribution to show the relative contribution of the different SN types at different radii, with respect to the distribution
of host galaxy \ha\ emission. 
This distribution is remarkably flat and consistent with the CC SNe being drawn from the same stellar population as that of the recent SF,
and the mean  
\textit{Fr}$_{\textit{\ha}}$ is 0.516 (0.023).
We can therefore conclude that
the \ha\ emission within galaxies accurately traces the CC SN progenitor parent stellar population, and this further
implies that both CC SNe and \ha\ emission are excellent tracers of SF within galaxies.
This plot 
then implies that (as discussed above) the central deficit of SNII (and the central excess of SNIb/c) is due to higher metallicities resulting in relatively more SNIb/c than SNII,
while as one goes further out into the disk of galaxies the lower metallicities inhibit the production of SNIb/c relative to SNII.\\
%The most logical interpretation of this trend
%is that at higher metallicity the minimum initial mass for producing SNIb/c lowers quite considerably and 
%therefore the production of SNIb/c is favoured over SNII. 
%An important point to note here is that in the central parts of galaxies this plot implies that less SNII
%are produced in the central parts of galaxies than would be expected by the amount of SF that is ongoing in these regions. One should
%also not that this cannot simply be put down to SNe exploding as SNIb/c \textit{instead} of SNII due to higher metallicites,
%because the shape of the IMF means that this would not be able to account for the significant drop in the numbers of SNII
%in the central regions.
%The distribution of the overall CC SN population with respect to the $R$-band light (shown in Fig.~\ref{figCCR})
%is also statistically flat and the mean \textit{Fr}$_{\textit{R}}$  is 0.519 (0.021). Galaxies are often thought of as having an old stellar bulge in their
%centres and then a younger outer disk. However this plot shows that the same fraction of CC SNe are being produced 
%equally at all positions within the $R$-band light. This suggests that the SFR per unit stellar mass is equal
%in both the galaxy bulge centres and out in the disk. This is consistent with the recent findings of James et al. (2008, A\&A submitted) who
%observe that the radial distributions of the $R$-band light and \ha\ emission of galaxies within the \ha GS survey
%agree remarkably well.\\
Another explanation for the above results is that there are deviations from a `normal' IMF as a function of radial
positions. \cite{man07} investigated the properties of 329 late-type giant stars in the central parts of the Galaxy. They concluded that
these stars originated from a top-heavy IMF. If this applies to the current sample
then the above results can be explained by relatively more high mass stars being formed and hence fractionally more
progenitor stars exploding as SNIb/c than as SNII when compared to the outer parts of galaxies with standard IMF 
stellar populations. However, this would not naturally produce the flat distribution seen in Fig.~\ref{figCCHa}
(in fact, the higher fraction of high mass stars would give more central \ha\ emission per SN progenitor star).\\
It can be asked whether the 
fraction of different events within the current sample is a true reflection of reality as the sample was
not chosen to be complete. However when the make up of the sample is compared to recent
local estimates of the relative SN rates (e.g. \citealt{sma09}), good agreement is found (the
only real discrepancy is the additional numbers of the SNIIL, IIb and IIn at the expense of SNIIP).
The percentages of SNII and SNIb/c in the current sample are 66\%\ and 34\%\ respectively compared to the Smartt et al. 
estimates of 71\%\ and 29\%.\\
The different CC SNe discussed above have different characteristic absolute magnitudes. This could therefore 
affect their relative detections at different radial positions (e.g. there may be more extinction or a higher
stellar background in central regions). \cite{rich02} used the Padova-Asiago SN catalogue to carry out a comparative study of SN
absolute magnitude distributions. They determined mean peak absolute magnitudes in the $B$-band of --18.04 for SNIb/c, --17.00
for SNIIP, --18.03 for SNIIL and finally --19.15 for SNIIn. One may speculate that the fainter magnitudes of
SNIIP (the most abundant SNII sub-type) compared to the SNIb/c naturally explains the higher fraction of SNIb/c 
in the centres of galaxies where there is likely to be higher extinction and a higher stellar background, preventing the detection of lower
luminosity events (however the SNIIP have a plateau phase in their light curves after maximum light meaning that they will
be detectable over a longer time period).
It is possible that these selection effects could affect the above results, however this effect would have to be quite severe and 
it is interesting to note that there are no SNIIn (out of 12 in the current sample),
the brightest CC SNe according to \cite{rich02}, detected within the central 20\%\ of the SF or the central 30\%\ of the $R$-band light. \\

It is stressed here that while there are many processes that may be at play in producing more or less 
of the different CC SNe at different radii, nearly all those discussed above would not produce a flat distribution
as is seen in Fig.~\ref{figCCHa}. The only interpretation of these results that naturally produces the 
extremely flat distribution observed, is that CC SNe are produced from a constant mass range of stars that have a non-changing
IMF, and that the change in relative numbers of SNII and SNIb/c seen in the centres of galaxies is down to there
being a substantial metallicity dependence of progenitors. Within the central 10\%\ of the SF within host galaxies
more SNIb/c than SNII have been produced within the current sample. 
%If this is simply down to progenitor
%metallicity then the dependence of SN type on parent stellar population metallicity is substantially
%stronger than has been suggested elsewhere.

\subsection{SNII sub-types}
\label{IIsub}
In \S ~\ref{compareIIP} the distributions of the SNIIP were compared to those of all the other SNII sub-types and
the SN `impostors' (the small number of SNe of each of the other sub-types meant that a statistical
comparison of each individually was not viable). It can be seen that all the other sub-types tend to
preferentially occur at larger radii with respect to both the $R$-band light and the \ha\ emission (see Figs.~\ref{figradIIPR},
~\ref{figradIIPHa},~\ref{nonPR} and ~\ref{nonPHa} plus the mean values given for the various types in \S ~\ref{results}). While there
only seems a slight deficit in SNIIP at small radii with respect to the $R$-band light, there is only one
SN of the other types that has exploded within the inner 30\%\ of the light (this difference is also seen with respect 
to the \ha\ emission, but to a lesser extent). Again if we regard these differences in terms of radial metallicity
gradients within host galaxies then one comes to the conclusion that SNIIP arise from more metal rich progenitors than 
the other sub-types. However, the spectra of the SNII sub-types indicate that their progenitors can be 
sorted into a sequence of increasing pre-SN stellar mass-loss going from SNIIP to IIL, IIb and finally IIn
(\citealt{chev06}). Therefore if there was a metallicity dependence
on producing the SNIIP compared to the other sub-types one would presume that in fact SNIIP would be found in less
metal rich environments than those of the SNIIL, IIb, IIn and SN `impostors' (higher metallicity progenitors
are likely to have stronger stellar winds). Theory also predicts that SNIIL and IIb will only be produced 
above some limiting metallicity (\citealt{heg03}).
While these differences between SNII sub-type environments
are not statistically significant, it is clear that there is some process that is preventing SNIIL, IIb, IIn and
SN `impostors' from being detected or occurring within the central parts of galaxies. The selection effects that may
be at play to reduce the detections of these events in the central regions of galaxies were discussed
in the previous section, and while these may seem valid by themselves the detection of large numbers of SNIb/c 
in the same central regions again makes these explanations unsatisfactory. 
%(although it is probable that there is a strong
%selection effect against detecting the low luminosity `impostor' events in the central parts of galaxies). 
The most
plausible explanation seems therefore to be that there is some higher metallicity limit that inhibits the production
of these SNe.

\subsection{Morphological classifications of SN host galaxies}
\label{galclass}
One may question whether there are intrinsic differences between the host galaxies of the different SN types that may influence our
results. To investigate this possibility morphological classifications were taken from the Padova-Asiago SN catalogue
for each SN host galaxy in the current sample. There were 9 galaxies within our sample for which classifications were not available. For these we 
used our own $R$-band imaging to assign morphological types. All host galaxy classifications were then changed to T-type classifications following the 
prescription in \cite{deva59}. The sample was then split into SNII, SNIb and SNIc host galaxies. Mean T-types were 
found of; 4.60 (standard error on the mean of 0.18) for the SNII, 4.73 (0.39) for SNIb and 4.45 (0.37) for SNIc, and
using a KS test it was found that all three galaxy distributions were consistent with being drawn from the same parent population. 
It therefore seems that there are no significant differences in host galaxy characteristics that could strongly
bias the results presented above. It is interesting to note that the distribution of T-type classifications for the galaxies in the current sample
peaks between Sbc and Sc galaxies, consistent with the findings of
\cite{jame08}, that the SFR in the local universe is concentrated
within these galaxy types.\\

\subsection{The radial distribution of SNe with respect to host galaxy D$_{25}$ and the
nature of galaxy metallicity gradients}
\label{d25}
The analysis above was based on fractions of total galaxy fluxes lying within SN locations.
However, metallicity gradients in galaxies are usually quoted as logarithmic
changes over a radial range normalised to an isophotal galaxy size, e.g. dex per (r/D$_{25}$).
Therefore, here we present a further analysis using r/D$_{25}$ in place of fractional luminosity.
%The arguments above on the implications for progenitor metallicity from differences in SN radial positions 
%have used an analysis based on galactocentric radial positions, normalising the fluxes contained within that position 
%to the total flux of the host galaxies. This has enabled discussion of the distribution of CC SNe with respect to different stellar
%populations and in particular how the CC SN population is distributed with respect to the SF within galaxies (see Fig.~\ref{figCCHa}).
%However, metallicity gradients within galaxies are usually quoted in some quantity such as change in metallicity in dex 
%per unit distance normalised to galaxy D$_{25}$ isophotal
%diameter. Therefore, here we present a further analysis of the
%distribution of
%galactocentric radial positions of SNe in the current sample, in terms of distance normalised to host galaxy D$_{25}$.
We do this to a) check the robustness of the above results and discussion, and b) to make some quantitative
statements on the differences in metallicity of the host stellar environments of CC SNe.\\
For the complete SN and host galaxy sample discussed above we used the radius of the semi-major axis (in arcsec) of the aperture that just
includes the SN position (calculated in \S ~\ref{rad}) and normalise this by dividing by the D$_{25}$ semi-major axis
of the host galaxy taken from NED, giving a value 
of r/D$_{25}$. The distribution of these values for the SNII, SNIb and SNIc populations is shown in Fig.~\ref{figd25}.
\begin{figure}\centering
\includegraphics[width=8cm]{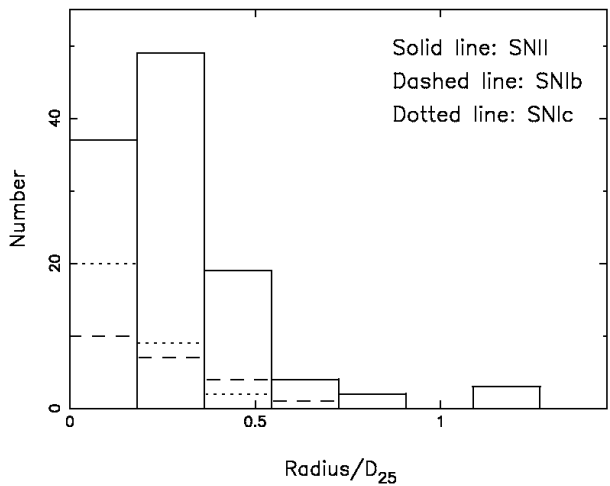}
\caption{Histogram of the galactocentric radial positions of the
  different CC SNe types normalised to their host galaxy D$_{25}$
  isophotal diameters. The solid line shows the distribution of SNII,
  dashed line the SNIb, and the dotted line the SNIc}
\label{figd25}
\end{figure}
This plot shows qualitatively the same results as shown in the histograms presented in 
\S ~\ref{results} and the cumulative distribution plot shown in Fig.~\ref{radmetR}. 
The SNIc are the most centrally concentrated distribution, followed by the SNIb and the SNII,
implying, as above, a decreasing sequence of progenitor metallicity.
The mean r/D$_{25}$ for the SNIc is 0.149 (standard error on the mean of 0.024), 0.244 (0.032) for the SNIb and 0.295 (0.020) for the 
SNII distribution. Using KS tests we find that there is less than 0.1\% probability that the SNIc are drawn from the same radial distribution as that of the
SNII, while the SNIb are consistent with arising from the same parent distribution as that of the SNII. There is a 7.7\% chance that the SNIb and SNIc 
distributions arise from the same parent population.\\
\cite{hen99} published metallicity gradients for a range of galaxy types. They found a mean metallicity gradient within galaxies  
of [O/H]= --1.0(r/D$_{25}$) (a value also observed by \citealt{pran00}). Using this gradient together with the differences between the mean values 
of r/D$_{25}$ presented above we estimate that the mean metallicity difference between the stellar environments of SNIc and
SNII is roughly 0.15 dex, i.e. that SNIc arise from progenitors with
metal abundances a factor of 1.4 higher on average than those of SNII.\\
The above analysis shows that there are real differences in the
radial positions of the different CC SN types. However, it is not clear whether the
metallicity differences that we derive can alone produce the differences in the relative rates of SNIc and SNII
that we find between galaxy centres and out in the disc, and therefore other factors must be at play. 
As discussed earlier, studies investigating the 
global host galaxy properties \citep{pran03,pri08_2} have concluded that 
SNIb/c are generally found in higher metallicity galaxies than SNII. We therefore investigated this in 
the current sample. Host galaxy absolute $B$-band magnitudes were calculated using galaxy
photometry and distances taken from NED. Surprisingly we do not recover the differences in host galaxy 
magnitude between the SNIb/c and SNII published by \citeauthor{pran03}, and all host galaxy magnitude distributions are 
statistically the same. It is not clear what is the cause of this discrepancy because, as shown 
above the host galaxy morphologies in our sample are typical for SF galaxies and therefore one would assume typical 
for the \citeauthor{pran03} sample, which was drawn from the Padova-Asiago SN catalogue. One possibility is that
the current sample is too small to observe differences in the host galaxy magnitudes (and we also note the size of the error bars in their analysis).
We also note that in interpreting these results, it should be borne in mind that many SNe are detected
in searches targeted on bright galaxies, which may somewhat affect the metallicity effects between galaxies, although any effect can only be small.\\
While \citeauthor{pran03} used host galaxy magnitude as a proxy for galaxy metallicity, \cite{pri08_2} obtained host galaxy
metallicities derived from spectra of the host galaxies and again found that SNIb/c were typically found in more metal rich galaxies than SNII.
Therefore it seems likely that SNIb/c arise from both more metal rich galaxies and, as found 
in the current study more metal rich environments within those galaxies. 
However, we concur with the conclusion of \citeauthor{pri08_2}, that metallicities should be 
directly derived at the SN locations, and we have such a project underway (see also \citealt{mod08}, for a similar
study).
%However,
%due to the discrepancies above we comment (as \citeauthor{pri08_2} also did) that to obtain accurate metallicities for the
%different SN types metallicities must be directly derived at the sites where SNe are found within galaxies,
%as has been achieved by \cite{mod08}, and as we will present in future publications.

\subsection{Implications for progenitor metallicity or mass?}
\label{massZ}
In this paper we have generally assumed that
the increased centralisation of a SN distribution implies increased 
progenitor metallicity. Recently \cite{kel08} presented statistics on the association of
SNe to the \sdssg-band light of their host galaxies (using a similar fractional flux technique
as that presented in AJ08), which has been modeled by \cite{ras08} to derive initial
progenitor mass estimates. 
This work showed that SNIc are concentrated towards 
the brightest parts of their host galaxies while the SNIb and SNII populations followed
the distribution of \sdssg-band light. These authors then argue that the 
brightest regions of galaxies are likely to correspond to the largest SF regions,
where the most massive stars are formed, a plausible argument given that SF results in local enhancement of \sdssg-band
luminosity. Using this assumption they conclude that SNIc arise from higher mass progenitors than SNII and SNIb.
However, this interpretation is ambiguous, since 
\sdssg-band brightness can be enhanced from other causes, and in particular shows a systematic trend with radius, and 
hence plausibly with metallicity, which complicates the interpretation of their result.\\
Our method, presented in AJ08 and the current paper, attempts to disentangle effects due to progenitor mass/age 
and those due to metallicity. The correlations of AJ08 make use of the continuum-subtracted \ha\ surface brightness,
which explicitly traces only the young stellar population, independently of radial position and local metallicity.
The present paper demonstrates that, within that very young stellar population, the part laying closest 
to the galaxy centres tends to produce large numbers of SNIc, and fewer SNII. These radial changes we 
attribute to metallicity effects. 
%that there are more stars at that location and not necessarily a larger
%fraction of high mass stars. The statistics presented here offer a different explanation to these
%results. The brightest regions of \sdssg-band flux are likely to be found towards the centers
%of galaxies (even if one removes the bulge population). Therefore it may be the progenitor
%metallicity (rather than mass) effect that we propose here, that is producing the distributions that they observe.\\
%As discussed earlier, in AJ08 we presented statistics using fractional \textit{\ha} fluxes to
%imply that SNIb/c arise from higher mass progenitors than SNII. While we understand that these \textit{conclusions}
%are similar to those of \cite{kel08}, we are not convinced that both studies are observing the same effect. 
%Overall we believe that both progenitor metallicity and mass produce the spatial correlations with
%host galaxy characteristics that are seen in nature. The degree to which these effects are correlated remains
%to be seen and will be investigated in future publications.

\subsection{Progenitor binarity}
\label{binary}
One major point of discussion that we have up to this point ignored
is the possibility that a large fraction of CC SN progenitors (in
particular the SNIb/c) may arise from binary systems. As discussed
earlier the relative
ratio of SNIb/c to SNII when compared with predictions from single star models and observations of massive stars in 
the local group (\citealt{mass03}, \citealt{crow07}), would seem to suggest that the production of a significant fraction of SNIb/c from binaries
is required. 
While a detailed discussion of CC SN production through binaries and how 
this may be affected by environment is beyond the scope of this paper, here
we briefly summarise the points that may be relevant to the current work.
As with single stars, theory in general predicts that the production of
SNIb/c from binaries should increase with increasing progenitor metallicity (see
\citealt{fry07} and \citealt{eld08}, although also see \citealt{izz04} for differing predictions). If 
correct, the higher degree of centralisation of SNIb/c (and in particular SNIc), 
can be explained again in terms of increasing metallicity, which increases the relative
production of these SNe through both single \textit{and} binary progenitor scenarios.
We also note that \cite{eld08} have shown that the observed trends
of the ratio of SNII to SNIb/c with metallicity (they use the SN host galaxy
luminosity observations of \citealt{pran03}) are most convincingly reproduced from
binary model predictions. Therefore, while
in general we have discussed our results in terms of single star SN 
progenitors, it seems that the implications of our results are
also compatible with current thinking on SN production from
binary channels, and further work is needed in order to
further constrain the fraction of events produced by the different
progenitor scenarios.

%It may be speculated that within different environments the
%fraction of binary stars may vary and this could be
%affecting the radial distribution of SNe that we are observing,
%however further investigation of this possibility is beyond the
%scope of this work and it remains to be seen whether the binary
%fraction of stars is dependent on the stellar environment in which
%SF takes place. Therefore while binarity may have a role in determining SN type,
%at present we would conclude that it does not naturally explain the radial
%distribution patterns found here.

\section{Conclusions}
\label{con}
We have presented results and discussion on the radial distribution of
CC SNe with respect to both old stellar populations as traced by the
$R$-band continuum light and young stellar populations as traced by
the \ha\ emission within host galaxies. We find that the distribution
of the overall CC SN population closely follows the distribution of
\ha\ line emission, implying that both are accurate tracers of SF
within galaxies. Within this correlation we find that the ratio of SNII to
SNIb/c is strongly correlated with galactocentric radial position, with
the SNIb/c (in particular the SNIc) dominating the CC SN production
within the central parts of the host galaxies' SF while as one moves out
into the disk the SNII become more dominant. This 
implies a strong metallicity dependence on
the relative production of these SN types from massive star
progenitors. We now list our main conclusions that arise from this work:
\begin{itemize}
\item
The radial distribution of CC SNe closely follows the radial distribution of
\ha\ emission within galaxies with a central deficit in the number of SNII offset
by a central excess of SNIb/c. This implies that both CC SNe and \ha\ emission
are excellent tracers of SF within galaxies
\item
SNIb/c are observed to occur within more central positions of host
galaxies than SNII implying that they arise from more metal rich
progenitors. This confirms and strengthens the results of \cite{bergh97}, \cite{tsv04}
and \cite{hak08}
\item
SNIc are seen to be the most centrally concentrated of all CC SN types
with
many SNe occurring within the central 20\%\ of the host galaxy light
and none found to occur in the outer 20\% . This implies that the majority of SNIc
occur within metal rich environments
\item
SNIb are found to trace the host galaxy light and line emission more
accurately than the SNIc while still being more centrally concentrated
than the SNII, implying a progenitor metallicity sequence going through SNII, SNIb and
SNIc, with the latter arising from the highest metallicity progenitors 
\item
SNIIP are found to be more centrally concentrated than the other SNII
sub-types
implying that SNe of types SNIIL, IIb, and IIn generally arise from
less metal rich progenitors than SNIIP, in apparent contradiction to theoretical 
predictions
\end{itemize}

\section*{Acknowledgments}
We thank the anonymous referee for their constructive comments.
We also thank Stephen Smartt and Mike Bode for useful discussion, and Mike Irwin 
for processing the INT data through the CASU WFC automated reduction pipeline. This research
has made use of the NASA/IPAC Extragalactic Database (NED) which is operated by the Jet Propulsion Laboratory, California
Institute of Technology, under contract with the National Aeronautics and Space Administration.

\bibliographystyle{mn2e}

\bibliography{Reference}

\appendix

\section[]{SN and host galaxy data}

\begin{table*}
\label{SNtabrad} 
\centering
\tablename{ A1. Data for all SNe and host galaxies}
\begin{tabular}[t]{ccccccccc}
\hline
\hline
SN & SN type & Host galaxy& Galaxy type &V$_\textit{r}$ (\kms ) & \textit{Fr}$_{\textit{R}}$ & \textit{Fr}$_{\textit{\ha}}$ & Filter & Reference\\
\hline

 1921B &  II& NGC 3184& SABcd&592&  0.856&  0.954&$R$&\\
 1926A & IIL& NGC 4303& SABbc& 1566&  0.607&  0.736&$R$&\\
 1937F & IIP& NGC 3184& SABcd&592&  0.808&  0.930&$R$&\\
 1940B & IIP& NGC 4725& SABab&1206&  0.675&  0.802&$R$&\\
 1941A & IIL& NGC 4559& SABcd&816&  0.208&  0.131&$R$&\\
 1941C &  II& NGC 4136& SABc&609&  0.880&  0.882&$R$&\\
 1954A &  Ib& NGC 4214& IABm&291&  0.968&  0.992&$R$&\\
 1954C &  II& NGC 5879& SAbc&772&  0.615&  0.511&$R$&\\
 1961I &  II& NGC 4303& SABbc&1566&  0.697&  0.877&$R$&\\
 1961V & `impostor$^{*}$'& NGC 1058&SAc &518&  0.968&  0.931&$R$& \cite{good89}\\
 1962L &  Ic& NGC 1035& SAc&1208&  0.754&  0.518&\textit{r'}&\\
 1964A &  II& NGC 3631& SAc&1156&  0.915&  0.992&$R$&\\
 1964F &  II& NGC 4303& SABbc&1566&  0.189&  0.106&$R$&\\
 1964H &  II& NGC 7292& IBm&986&  0.551&  0.719&$R$&\\
 1965H & IIP& NGC 4666& SABc&1529&  0.324&  0.198&\textit{r'}&\\
 1965L & IIP& NGC 3631& SAc&1156&  0.622&  0.658&$R$&\\
 1965N & IIP& NGC 3074& SABc&5144&  0.110&  0.059&$R$&\\
 1966B & IIL& NGC 4688& SBcd&986&  0.571&  0.454&\textit{r'}\\
 1966J &  Ib& NGC 3198& SBc&663&  0.898&  0.936&$R$&\\
 1967H &  II$^{*}$& NGC 4254& SAc&2407&  0.664&  0.648&$R$& \cite{vandyk92}\\
 1968V &  II& NGC 2276& SABc&2410&  0.699&  0.790&$R$&\\
 1969B & IIP& NGC 3556& SBcd&699&  0.197&  0.494&$R$&\\
 1969L & IIP& NGC 1058& SAc&518&  1.000&  1.000&$R$&\\
 1971K & IIP& NGC 3811& SBcd&3105&  0.809&  0.900&$R$&\\
 1971S & IIP&  NGC 493& SABcd&2338&  0.605&  0.570&$R$&\\
 1972Q & IIP& NGC 4254& SAc&2407&  0.811&  0.791&$R$&\\
 1972R &  Ib& NGC 2841& SAb&638&  0.855&  0.904&$R$&\\
 1973R & IIP& NGC 3627& SABb&727&  0.471&  0.566&$R$&\\
 1975T & IIP& NGC 3756& SABbc&1318&  0.846&  0.856&$R$&\\
 1982F & IIP& NGC 4490& SBd&565&  0.277&  0.202&$R$&\\
 1983I &  Ic& NGC 4051& SABbc&700&  0.498&  0.473&$R$&\\
 1984E & IIL& NGC 3169& SAa&1238&  0.684&  0.731&$R$&\\
 1985F &  Ib$^{*}$& NGC 4618& SBm&544&  0.121&  0.087&\textit{r'}& \cite{gas86} \\
 1985G & IIP& NGC 4451& Sbc&864&  0.138&  0.212&$R$&\\
 1985L & IIL& NGC 5033& SAc&875&  0.585&  0.571&$R$&\\
 1986I & IIP& NGC 4254& SAc&2407&  0.334&  0.318&$R$&\\
 1987F & IIn& NGC 4615& Scd&4716&  0.489&  0.333&$R$&\\
 1987K & IIb& NGC 4651& SAc&805&  0.409&  0.303&$R$&\\
 1987M &  Ic& NGC 2715& SABc&1339&  0.129&  0.044&$R$&\\
 1988L &  Ib& NGC 5480& SAc&1856&  0.230&  0.369&\textit{r'}\\
 1989C & IIP& UGC 5249& SBd&1874&  0.017&  0.058&\textit{r'}\\
 1990E & IIP& NGC 1035& SAc&1241&  0.272&  0.363&\textit{r'}\\
 1990H & IIP$^{*}$& NGC 3294& SAc&1586&  0.156&  0.125&$R$& \cite{fil2_93}\\
 1990U &  Ic& NGC 7479& SBc&2381&  0.603&  0.488&$R$&\\
 1991A &  Ic&  IC 2973& SBd&3210&  0.742&  0.588&$R$&\\
 1991G & IIP& NGC 4088& SABbc&757&  0.466&  0.453&$R$&\\
 1991N &  Ic& NGC 3310& SABbc&993&  0.268&  0.277&$R$&\\
 1992C &  II& NGC 3367& SBc&3040&  0.689&  0.687&$R$&\\
 1993G & IIL$^{*}$& NGC 3690& Double&3121&  0.464&  0.744&$R$&\cite{tsvet94} \\
 1993X &  II& NGC 2276& SABc&2410&  0.899&  0.619&$R$&\\
 1994I &  Ic& NGC 5194& SAbc&463&  &  0.122&$R$&\\
 1994Y & IIn& NGC 5371& SABbc&2558&  0.355&  0.212&$R$&\\
1994ak & IIn& NGC 2782& SABa&2543&  0.725&  0.977&\textit{r'}\\
 1995F &  Ic& NGC 2726& SABc&2410&  0.037&  0.050&$R$&\\
 1995N & IIn& MCG -02-38-17&IBm &1856&  0.612&  0.822&\textit{r'}\\
 1995V &  II& NGC 1087& SABc&1517&  0.368&  0.497&$R$&\\
1995ag &  II& UGC 11861& SABdm&1481&  0.343&  0.170&$R$&\\
1996ae & IIn& NGC 5775& Sb&1681&  0.757&  0.671&$R$&\\
1996ak &  II& NGC 5021& SBb&8487&  0.619&  0.659&$R$&\\
1996aq &  Ic& NGC 5584& SABcd&1638&  0.178&  0.086&\textit{r'}\\
1996bu & IIn& NGC 3631& SAc&1156&  0.923&  0.993&$R$&\\
1996cc &  II& NGC 5673& SBc&2082&  0.924&  0.934&$R$&\\
 1997X &  Ic& NGC 4691& SB0/a&1110&  0.171&  0.472&$R$&\\	
1997bs & `impostor$^{*}$'& NGC 3627& SABb&727&  0.362&  0.348 &$R$&  \cite{van00}\\

\end{tabular}
\end{table*}

\setcounter{table}{0}
\begin{table*}
\centering
\tablename{ A1. Data for all SNe and host galaxies}
\begin{tabular}[t]{ccccccccc}
\hline
\hline
SN & SN type & Host galaxy& Galaxy type& V$_\textit{r}$ (\kms ) & \textit{Fr}$_{\textit{R}}$ & \textit{Fr}$_{\textit{\ha}}$ & Reference\\
\hline
1997db &  II& UGC 11861& SABdm&1481&  0.633&  0.396 &$R$&\\
1997dn &  II& NGC 3451& Sd&1334&  0.872&  0.946 &$R$&\\
1997dq &  Ic$^{*}$& NGC 3810& SAc&993&  0.774&  0.734 &$R$& \cite{mazz04}\\
1997eg & IIn& NGC 5012& SABc&2619&  0.503&  0.449 &$R$&\\
1997ei &  Ic& NGC 3963& SABbc&3188&  0.197&  0.053 &$R$&\\
 1998C &  II& UGC 3825& SABbc&8281&  0.544&  0.377 &$R$&\\
 1998T &  Ib& NGC 3690& Double&3121&  0.056&  0.056 &$R$&\\
 1998Y &  II& NGC 2415& Im&3784&  0.612&  0.720 &$R$&\\
 1999D &  II& NGC 3690& Double&3121&  0.560&  0.849 &$R$&\\
1999br & IIP$^{*}$& NGC 4900& SBc&960&  0.786&  0.932 &$R$&\cite{ham03}\\
1999bu &  Ic& NGC 3786& SABa&2678&  0.180&  0.522 &$R$&\\
1999bw & `impostor$^{*}$'& NGC 3198&SBc &663&  0.745&  0.755&$R$&\cite{vandyk05}\\
1999dn &  Ib& NGC 3451& Sd&2798&  0.872&  0.946 &$R$&\\
1999ec &  Ib& NGC 2207& SABbc&2741&  &  0.521 &$R$&\\
1999el & IIn& NGC 6951& SABbc&1424&  0.320&  0.259 &$R$&\\
1999em & IIP& NGC 1637& SABc&717&  0.276&  0.268 &\textit{r'}\\
1999gb & IIn& NGC 2532& SABc&5260&  0.485&  0.443 &$R$&\\
1999gi & IIP& NGC 3184& SABcd&592&  0.276&  0.112 &$R$&\\
1999gn & IIP$^{*}$& NGC 4303& SABbc&1566&  0.418&  0.429 &$R$&\cite{past04}\\
 2000C &  Ic& NGC 2415& Im&3784&  0.706&  0.820 &$R$&\\
2000cr &  Ic& NGC 5395& SAb&3491&  0.538&  0.549 &$R$&\\
2000db &  II& NGC 3949& SAbc&800&  0.364&  0.253 &$R$&\\
2000de &  Ib& NGC 4384& Sa&2513&  0.087&  0.140 &$R$&\\
2000ew &  Ic& NGC 3810& SAc&993&  0.261&  0.147 &$R$&\\
 2001B &  Ib&   IC 391& SAc&1556&  0.062&  0.060 &$R$&\\
 2001M &  Ic& NGC 3240& SABb&3550&  0.323&  0.251 &$R$&\\
 2001X & IIP& NGC 5921& SBbc&1480&  0.579&  0.369 &$R$&\\
2001aa &  II& UGC 10888& SBb&6149&  0.830&  0.796 &$R$&\\
2001ac & `impostor$^{*}$'& NGC 3504& SABab&1534&  0.826&  0.992 &$R$& \cite{math01}\\
2001ai &  Ic& NGC 5278& SAb&7541&  0.323&  0.256 &$R$&\\
2001co & Ib/c& NGC 5559& SBb&5166&  0.618&  0.497 &$R$&\\
2001ef &  Ic&   IC 381& SABbc&2476&  0.082&  0.052 &$R$&\\
2001ej &  Ib& UGC 3829& Sb&4031&  0.152&  0.391 &$R$&\\
2001fv & IIP$^{*}$& NGC 3512& SABc&1376&  0.669&  0.689 &$R$&\cite{math2_01}\\
2001gd & IIb& NGC 5033& SAc&875&  0.692&  0.675 &$R$&\\
2001is &  Ib& NGC 1961& SABc&3934&  &  0.749 &$R$&\\
 2002A & IIn& UGC 3804& Scd&2887&  0.419&  0.253 &$R$&\\
2002bu & IIn& NGC 4242& SABdm&506&  0.896&  0.930 &$R$&\\
2002ce &  II& NGC 2604& SBcd&2078&  0.381&  0.560 &$R$&\\
2002cg &  Ic& UGC 10415& SABb&9574&  0.151&  0.092 &$R$&\\
2002cp & Ib/c& NGC 3074& SABc&5144&  0.936&  0.961 &$R$&\\
2002cw &  Ib& NGC 6700& SBc&4588&  &  0.642 &$R$&\\
2002dw &  II& UGC 11376& Sbc&6528&  0.431&  0.644 &$R$&\\
2002ed & IIP& NGC 5468& SABcd&2842&  0.811&  0.791 &$R$&\\
2002ei & IIP& MCG -01-09-24& Sab&2319&  0.195&  0.195 &\textit{r'}&\\
2002gd &  II& NGC 7537& SAbc&2674&  0.759&  0.685 &$R$&\\
2002hn &  Ic& NGC 2532& SABc&48&  0.023&  0.011 &$R$&\\
2002ho &  Ic& NGC 4210& SBb&2732&  0.146&  0.051&$R$&\\
2002ji & Ib/c& NGC 3655& SAc&1473&  0.709&  0.957&$R$&\\
2002jz &  Ic& UGC 2984& SBdm&1543&  0.091&  0.099&$R$&\\
 2003H &  Ib& NGC 2207& SABbc&2741&  &  0.259&$R$&\\
 2003T &  II& UGC 4864& SAab&8368&  0.682&  0.643&$R$&\\
 2003Z & IIP$^{*}$& NGC 2742& SAc&1289&  0.675&  0.736&$R$&\cite{past04}\\
2003ab &  II& UGC 4930& Scd&8750&  0.559&  0.454&$R$&\\
2003ao & IIP& NGC 2993& Sa&2430&  0.456&  0.784&\textit{r'}&\\
2003at &  II& MCG +11-20-33&Sbc &7195&  0.732&  0.871&$R$&\\
2003bp &  Ib& NGC 2596& Sb&5938&  0.486&  0.362&$R$&\\
2003db &  II& MCG +05-23-21&Sc &8113&  0.603&  0.603&$R$&\\
2003ed & IIb& NGC 5303& Pec&1419&  0.369&  0.298&\textit{r'}&\\
2003ef &  II$^{*}$& NGC 4708& SAab&4166&  0.335&  0.352&$R$&\cite{gan03}\\
2003el &  Ic& NGC 5000& SBbc&5608&  0.482&  0.476&$R$&\\
2003hp &  Ic& UGC 10942& SBbc&6378&  0.660&  0.742&$R$&\\
2003hr &  II& NGC 2551& SA0/a&2344&  0.914&  1.000&$R$&\\
2003ie &  II& NGC 4051& SABbc&700&  0.838&  0.885&$R$&\\
\end{tabular}
\end{table*}

\setcounter{table}{0}
\begin{table*}
\centering
\tablename{ A1. Data for all SNe and host galaxies}
\begin{tabular}[t]{ccccccccc}
\hline
\hline
SN & SN type & Host galaxy& Galaxy type& V$_\textit{r}$ (\kms ) & \textit{Fr}$_{\textit{R}}$ & \textit{Fr}$_{\textit{\ha}}$ & Reference\\
\hline
2003ig &  Ic& UGC 2971& Sbc&5881&  0.176&  0.108&$R$&\\
 2004A & IIP$^{*}$& NGC 6207& SAc&852&  0.729&  0.660&$R$&\cite{hen06}\\
 2004C &  Ic& NGC 3683& SBc&1716&  0.532&  0.545&$R$&\\
 2004D &  II& UGC 6916& SBbc&6182&  0.457&  0.336&$R$&\\
 2004G &  II& NGC 5668& SAd&1582&  0.657&  0.595&$R$&\\
 2004T &  II& UGC 6038& Sb&6446&  0.507&  0.260&$R$&\\
 2004Z &  II& MCG +10-19-85&SB0/a &6933&  0.548&  0.333&$R$&\\
2004ak &  II& UGC 4436& Sbc&1124&  0.887&  0.882&$R$&\\
2004ao &  Ib& UGC 10862& SBc&1691&  &  0.215&$R$&\\
2004au &  II& MCG +04-42-2& Sb&7800&  0.330&  0.446&$R$&\\
2004bi & IIb& UGC 5894& Sab&6537&  0.875&  0.846&$R$&\\
2004bm &  Ic& NGC 3437& SABc&1283&  0.073&  0.076&$R$&\\
2004bn &  II& NGC 3441& Sbc&6554&  0.509&  0.491&$R$&\\
2004bs &  Ib& NGC 3323& SBc&5164&  0.191&  0.119&$R$&\\
2004dk &  Ib& NGC 6118& SAcd&1573&  0.673&  0.626&$R$&\\
2004dg & IIP$^{*}$& NGC 5806& SABb&1359&  0.484&  0.378&$R$& S. Smartt (2008, priv comm)\\
2004eb &  II& NGC 6387& Pec&8499&  0.413&  0.540&$R$&\\
2004ed &  II& NGC 6786& SBc&7500&  0.550&  0.624&$R$&\\
2004ep &  II&  IC 2152& SABab&1875&  0.461&  0.560&\textit{r'}&\\
2004es &  II& UGC 3825& SABbc&8281&  0.974&  0.904&$R$&\\
2004ez & IIP& NGC 3430& SABc&1586&  0.788&  0.833&\textit{r'}&\\
2004gj & IIb&   IC 701& SBdm&6142&  0.674&  0.652&$R$&\\
2004gq &  Ib& NGC 1832& SBbc&1939&  0.672&  0.328&\textit{r'}&\\
2004gr &  II$^{*}$ & NGC 3678&Sbc &7210&  0.553&  0.570&$R$&\cite{pugh04} \\
2004gt & Ib/c& NGC 4038& SBm&1642&  0.834&  0.991&\textit{r'}&\\
 2005D & IIb$^{*}$ & UGC 3856&Scd &8505&  0.999&  0.981&$R$&\cite{gra05}\\
 2005K &  II& NGC 2923& S0/a&6538&  0.743&  0.784&$R$&\\
 2005O &  Ib& NGC 3340& SBbc&5558&  0.322&  0.305&$R$&\\
 2005V & Ib/c& NGC 2146&SBab &893&  0.033&  0.091&$R$&\\
2005ad & IIP&  NGC 941& Sc&1608&  0.831&  0.864&$R$&S. Smartt (2008, priv comm)\\
2005az &  Ic$^{*}$& NGC 4961& SBcd&2535&  0.426&  &\textit{r'}&\cite{burk05}\\
2005ci &  II& NGC 5682& SBb&2273&  0.204&  0.191&\textit{r'}&\\
2005cs & IIP& NGC 5194& SAbc&463&  &  0.222&$R$&\\
2005dl &  II& NGC 2276& SBc&2410&  0.247&  0.099&$R$&\\
2005dp &  II& NGC 5630& Sdm&2655&  0.534&  0.590&\textit{r'}&\\
2005ip &  II& NGC 2906& Scd&2140&  0.399&  0.528&\textit{r'}&\\
2005kk &  II& NGC 3323& SBc&5164&  0.766&  0.875&\textit{r'}&\\
2005kl &  Ic& NGC 4369& SAa&1045&  0.271&  0.540&\textit{r'}&\\
2005lr &  Ic& ESO 492-G2& Sb&2590&  &  0.005&\textit{r'}&\\
 2006T & IIb& NGC 3054&SABbc &2426&  0.302&  &\textit{r'}&\\
2006am & IIn& NGC 5630&Sdm &2655&  0.604&  0.617&\textit{r'}&\\
2006gi &  Ib& NGC 3147& SAbc&2820&  0.984&  0.991&\textit{r'}&\\
2006jc & Ib/c& UGC 4904& SB&1670&  0.332&  0.525&\textit{r'}&\\
2006ov & IIP& NGC 4303& SABbc&1566&  0.418&  0.429&$R$&\\
 2007C &  Ib& NGC 4981& SABbc&1680&  0.320&  0.236&$R$&\\
2007am & II &NGC 3367 & SBc&3040& 0.302 & 0.314 &\textit{r'}&\\
2007fp & II &NGC 3340 & SBbc&5558& 0.170 & 0.125 &\textit{r'}&\\
2008ax & IIb& NGC 4490& SBd&565&  0.328&  0.284&$R$&\\
\hline
\end{tabular}
%\begin{tablenotes}
%\item[]\footnotesize{$^1$\cite{pugh04}; $^2$\cite{gra05}; $^3$\cite{jha06}}
%\end{tablenotes}
\caption{Data for all SNe and host galaxies used in the analysis presented in this paper. In the first two columns the SN names and types are given.
In columns 3, 4 and 5 the host galaxy names, morphological types and recession velocities are listed. Next the \textit{Fr}$_{\textit{R}}$ and \textit{Fr}$_{\textit{\ha}}$
values derived for each SN from the above analysis are presented, and in column 8 the filter used for the observations is listed ($R$-band 
Johnson/Bessell or \textit{r'}-band Sloan). In the final column a reference is given for those SNe type classifications where
classification was changed from that given in the Asiago catalogue and these are also marked with an asterisk}

\end{table*}

\label{lastpage}	    
			    
\end{document}